\documentclass[acmtog,nonacm]{acmart}
\AtBeginDocument{%
  \providecommand\BibTeX{{%
    \normalfont B\kern-0.5em{\scshape i\kern-0.25em b}\kern-0.8em\TeX}}}

\newbool{authorcomments}
\boolfalse{authorcomments} %
\ifbool{authorcomments}%
{
\newcommand{\NML}[1]{{\color{blue}{[Nico: #1]}}}

\newcommand{\NS}[1]{{\color{olive}{[Nick: #1]}}}

\newcommand{\ZW}[1]{{\color{magenta}{[Zian: #1]}}}
\newcommand{\RDL}[1]{{\color{cyan}{[Riccardo: #1]}}}

\newcommand{\AM}[1]{{\color{teal}{[Ashkan: #1]}}}

\newcommand{\GS}[1]{{\color{brown}{[Gav: #1]}}}

\newcommand{\ZG}[1]{{\color{violet}{[Zan: #1]}}}

\newcommand{\OP}[1]{{\color{purple}{[Or: #1]}}}
\newcommand{\OPC}[1]{{\color{purple}{#1}}}
\newcommand{\JM}[1]{{\color{orange}{[Janick: #1]}}}

\newcommand{\SF}[1]{{\color{teal}{[Sanja: #1]}}}

}
{
\newcommand{\NML}[1]{{}}

\newcommand{\NS}[1]{{}}

\newcommand{\ZW}[1]{{}}
\newcommand{\RDL}[1]{{}}

\newcommand{\AM}[1]{{}}

\newcommand{\GS}[1]{{}}

\newcommand{\ZG}[1]{{}}

\newcommand{\OP}[1]{{}}
\newcommand{\OPC}[1]{{#1}}
\newcommand{\JM}[1]{{}}

\newcommand{\SF}[1]{{}}

}

\newcommand{\revAdded}[1]{{#1}}
\newcommand{\revRemoved}[1]{{}}
\newcommand{\revReplaced}[2]{#2}

\newcommand{\RayOrigin}{\bm{o}}
\newcommand{\RayDirection}{\bm{d}}
\newcommand{\ParticleCenter}{\bm{\mu}}
\newcommand{\ParticleScale}{\bm{s}}
\newcommand{\ParticleRotation}{\bm{q}}
\newcommand{\ParticleRadiance}{\phi}
\newcommand{\ParticleOpacity}{\sigma}
\newcommand{\ParticleHarmonics}{\bm{\beta}}
\newcommand{\ParticleResponse}{\rho}
\newcommand{\ParticleCosineModulation}{\psi}
\newcommand{\AlphaResponse}{\alpha}
\newcommand{\Radiance}{\bm{L}}

\newcommand{\MinResponse}{\AlphaResponse_\textrm{min}}
\newcommand{\MinAlphaResponse}{\MinResponse}
\newcommand{\SceneTMin}{\tau_\textrm{SceneMin}}
\newcommand{\SceneTMax}{\tau_\textrm{SceneMax}}
\newcommand{\CurrT}{\tau_\textrm{curr}}

\newcommand{\ParticleTraceBatchSize}{k}
\newcommand{\THit}{\tau_\textrm{hit}}
\newcommand{\TMax}{\tau_\textrm{max}}
\newcommand{\Transmittance}{T}
\newcommand{\MinTransmittance}{\Transmittance_\textrm{min}}

\definecolor{gold}{RGB}{221, 196, 65}
\definecolor{silver}{RGB}{215, 215, 215}
\definecolor{bronze}{RGB}{126, 66, 5}
\definecolor{green_zoomin}{RGB}{42, 137, 0}
\definecolor{nsteps_blue}{rgb}{0.03137254901960784, 0.18823529411764706, 0.4196078431372549}

\definecolor{palette0}{RGB}{140, 179, 105}
\definecolor{palette1}{RGB}{244, 226, 133}
\definecolor{palette2}{RGB}{244, 162, 89}
\definecolor{palette3}{RGB}{91, 142, 125}
\definecolor{palette4}{RGB}{188, 75, 81}

\acmSubmissionID{536}

\usepackage{hyperref}
\usepackage{multirow}
\usepackage{enumitem}
\usepackage{amsmath}
\usepackage{amsfonts}
\usepackage{nicefrac}
\usepackage{mathtools}
\usepackage{wrapfig}
\usepackage{tikz}
\usepackage{xcolor}
\usepackage{colortbl}
\usepackage{adjustbox}
\usepackage{booktabs} %
\usepackage{tabularx}
\usepackage{xspace}
\usepackage{ulem}
\usepackage[group-separator={,}]{siunitx}
\usepackage[htt]{hyphenat}
\newcolumntype{Y}{>{\centering\arraybackslash}X}
\newcolumntype{C}[1]{>{\centering\arraybackslash}p{#1}}

\graphicspath{{figures/}}

\usepackage{amsmath,amsfonts,bm}

\def\1{\bm{1}}

\def\vc{{\bm{c}}}

\DeclareMathAlphabet{\mathsfit}{\encodingdefault}{\sfdefault}{m}{sl}
\SetMathAlphabet{\mathsfit}{bold}{\encodingdefault}{\sfdefault}{bx}{n}

\usepackage{custom_style} %


\setcopyright{rightsretained}
\acmJournal{TOG}
\acmYear{2024} \acmVolume{43} \acmNumber{6} \acmArticle{232} \acmMonth{12}\acmDOI{10.1145/3687934}

\begin{document}

\title{3D Gaussian Ray Tracing: Fast Tracing of Particle Scenes}

\author{Nicolas Moenne-Loccoz}
\orcid{TODO}
\authornote{Authors contributed equally.}
\affiliation{%
  \institution{NVIDIA}
   \city{Montreal}
   \country{Canada}
  }
\email{nicolasm@nvidia.com}

\author{Ashkan Mirzaei}
\orcid{TODO}
\authornotemark[1]
\affiliation{%
  \institution{NVIDIA}
   \city{Toronto}
   \country{Canada}
  }
\affiliation{%
  \institution{University of Toronto}
   \city{Toronto}
   \country{Canada}
  }
\email{ashkan@cs.toronto.edu}

\author{Or Perel}
\orcid{TODO}
\affiliation{%
  \institution{NVIDIA}
   \city{Tel Aviv}
   \country{Israel}
  }
\email{operel@nvidia.com}

\author{Riccardo de Lutio}
\orcid{TODO}
\affiliation{%
  \institution{NVIDIA}
   \city{Santa Clara}
   \country{USA}
  }
\email{rdelutio@nvidia.com}

\author{Janick Martinez Esturo}
\orcid{0000-0002-6907-5639}
\affiliation{%
  \institution{NVIDIA}
   \city{Munich}
   \country{Germany}
  }
\email{janickm@nvidia.com}

\author{Gavriel State}
\orcid{TODO}
\affiliation{%
  \institution{NVIDIA}
   \city{Toronto}
   \country{Canada}
  }
\email{gstate@nvidia.com}

\author{Sanja Fidler}
\orcid{0000-0003-1040-3260}
\affiliation{%
  \institution{NVIDIA}
   \city{Toronto}
   \country{Canada}
  }
\affiliation{%
  \institution{University of Toronto}
   \city{Toronto}
   \country{Canada}
  }
\affiliation{%
  \institution{Vector Institute}
   \city{Toronto}
   \country{Canada}
  }
\email{sfidler@nvidia.com}

\author{Nicholas Sharp}
\orcid{0000-0002-2130-3735}
\authornote{Authors contributed equally.}
\affiliation{%
  \institution{NVIDIA}
   \city{Seattle}
   \country{USA}
  }
\email{nsharp@nvidia.com}

\author{Zan Gojcic}
\orcid{0000-0001-6392-2158}
\authornotemark[2]
\affiliation{%
  \institution{NVIDIA}
   \city{Z\"urich}
   \country{Switzerland}
  }
\email{zgojcic@nvidia.com}

\renewcommand{\shortauthors}{Moenne-Loccoz, Mirzaei \etal{}}

\acmSubmissionID{536}



\begin{abstract}
Particle-based representations of radiance fields such as 3D Gaussian Splatting have found great success for reconstructing and re-rendering of complex scenes.
Most existing methods render particles via rasterization, projecting them to screen space tiles for processing in a sorted order.
This work instead considers ray tracing the particles, building a bounding volume hierarchy and casting a ray for each pixel using high-performance GPU ray tracing hardware.
To efficiently handle large numbers of semi-transparent particles, we describe a specialized rendering algorithm which encapsulates particles with bounding meshes to leverage fast ray-triangle intersections, and shades batches of intersections in depth-order. 
The benefits of ray tracing are well-known in computer graphics: processing incoherent rays for secondary lighting effects such as shadows and reflections, rendering from highly-distorted cameras common in robotics, stochastically sampling rays, and more.
With our renderer, this flexibility comes at little cost compared to rasterization. Experiments demonstrate the speed and accuracy of our approach, as well as several applications in computer graphics and vision.
We further propose related improvements to the basic Gaussian representation, including a simple use of generalized kernel functions which significantly reduces particle hit counts.
\end{abstract}

\begin{teaserfigure}
\centering
\vspace*{-3mm}
\includegraphics{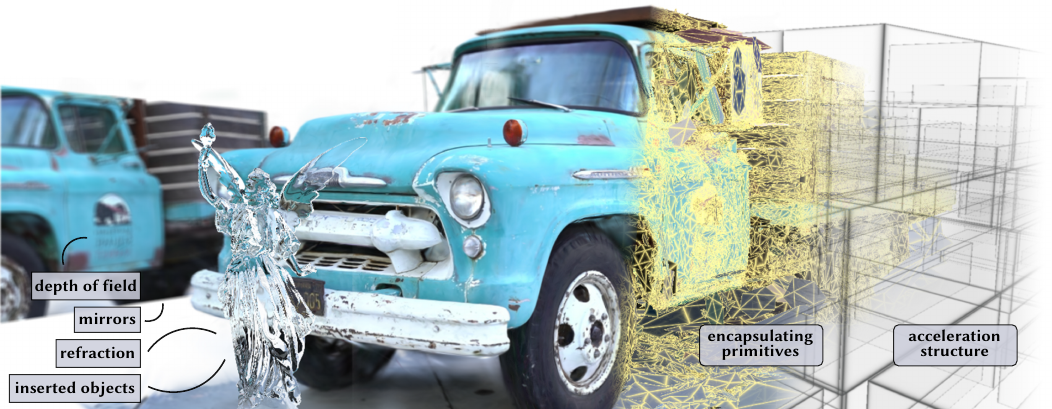}
\vspace{-1mm}
\caption{
We propose a method for fast forward and inverse ray tracing of particle-based scene representations such as Gaussians. 
The main idea is to construct encapsulating primitives around each particle, and insert them into a BVH to be rendered by a ray tracer specially adapted to the high density of overlapping particles.
Efficient ray tracing opens the door to many advanced techniques, including secondary ray effects like mirrors, refractions and shadows, as well as highly-distorted cameras with rolling shutter effects and even stochastic sampling of rays.
\bfseries Project page: \href{https://GaussianTracer.github.io}{GaussianTracer.github.io}
\label{fig:teaser}
\vspace{1em}
}
\end{teaserfigure}

\maketitle

\vfill

\pagebreak
\section{Introduction}

Multiview 3D reconstruction and novel-view synthesis are a classic challenge in visual computing, key to applications across robotics, telepresence, AR/VR, and beyond.
Many approaches have been proposed, but most recently particle-based representations have shown incredible success, ignited by 3D Gaussian Splatting~\cite{kerbl3Dgaussians} (3DGS)---the basic idea is to represent a scene as a large unstructured collection of fuzzy particles which can be differentiably rendered by splatting to a camera view with a tile-based rasterizer. The location, shape, and appearance of the particles are optimized using a re-rendering loss. 

Meanwhile, more broadly in computer graphics, rendering has long been a duality between rasterization and ray tracing.
Traditionally, rasterization supported real-time performance at the expense of approximating image formation, while ray tracing enabled fully general high-fidelity rendering in the expensive offline setting.
However, the introduction of specialized GPU ray tracing hardware and real-time renderers has moved ray tracing into the realm of real-time performance.

This work is motivated by the observation that 3DGS is limited by the classic tradeoffs of rasterization. The tile-based rasterizer is ill-suited to rendering from highly-distorted cameras with rolling shutter effects, which are important in robotics and simulation. 
It also cannot efficiently simulate secondary rays needed to handle phenomena like reflection, refraction, and shadows.
Likewise, rasterization cannot sample rays stochastically, a common practice for training in computer vision.
Indeed, prior work has already encountered the need for these capabilities, but was limited to rendering with restrictive tricks or workarounds~\cite{niemeyer2024radsplat, seiskari2024gaussian}.
We instead aim to address these limitations by making the ray traced Gaussian particles efficient, with a tailored implementation for specialized GPU ray tracing.
To be clear, goal of this work is not to offer an end-to-end solution to unsolved problems like global illumination or inverse lighting on particle scenes, but rather to provide a key algorithmic ingredient to future research on these problems: a fast differentiable ray tracer.

Efficiently ray tracing Gaussian scenes (and more generally semi-transparent surfaces) is not a solved problem \cite{TankiZhang2021}.
We find that even past algorithms that were specially designed for ray tracing semi-transparent particles~\cite{mboit2018,knoll2019,mlat2020} are ineffective on these scene reconstructions, due to the huge numbers of non-uniformly distributed and densely-overlapping particles.
Accordingly, we design a customized GPU-accelerated ray tracer for Gaussian particles with a $\ParticleTraceBatchSize$-buffer \cite{kbuffer2007} hits-based marching to gather ordered intersections, bounding mesh proxies to leverage fast ray-triangle intersections, and a backward pass to enable optimization.
Each of these components is carefully tested for speed and quality on a variety of benchmarks. 
We found it crucial to tune the details of the algorithm to the task at hand. Our final proposed algorithm is almost 25x faster than our first naive implementation, due to a wide range of algorithmic and numerical factors. We hope that these learnings will be of value to the community leveraging raytracing on particle representations. 

The fundamental approach of representing a scene with particles is not limited to the Gaussian kernel; and recent work has already shown several natural generalizations~\cite{Huang2DGS2024}.
Our ray tracing scheme, as well as the benefits and applications above, likewise generalizes more broadly to particle-based scene representations, as we show in section ~\secref{generalized_particles}.

We evaluate this approach on a wide variety of benchmarks and applications.
On standard multiview benchmarks, ray tracing nearly matches or exceeds the quality of the 3DGS rasterizer of \citet{kerbl3Dgaussians}, while still achieving real-time rendering framerates.
More importantly, we demonstrate a variety of new techniques made easy and efficient by ray tracing, including secondary ray effects like shadows and reflections, rendering from cameras with high distortion and rolling shutter, training with stochastically sampled rays and more.

In summary, the contributions of this work are:
\begin{itemize}[topsep=4pt, leftmargin=20pt]
\item A GPU-accelerated ray tracing algorithm for semi-transparent particles.

\item An improved optimization pipeline for ray-traced, particle-based radiance fields.
\item Generalized Gaussian particle formulations that reduce the number of intersections and lead to improved rendering efficiency.
\item Applications demonstrating the utility of ray tracing, including: depth of field, shadows, mirrors, highly-distorted cameras, rolling shutter, incoherent rays, and instancing.
\end{itemize}

\begin{figure*}[!t]
    \centering
    \includegraphics[width=1\textwidth]{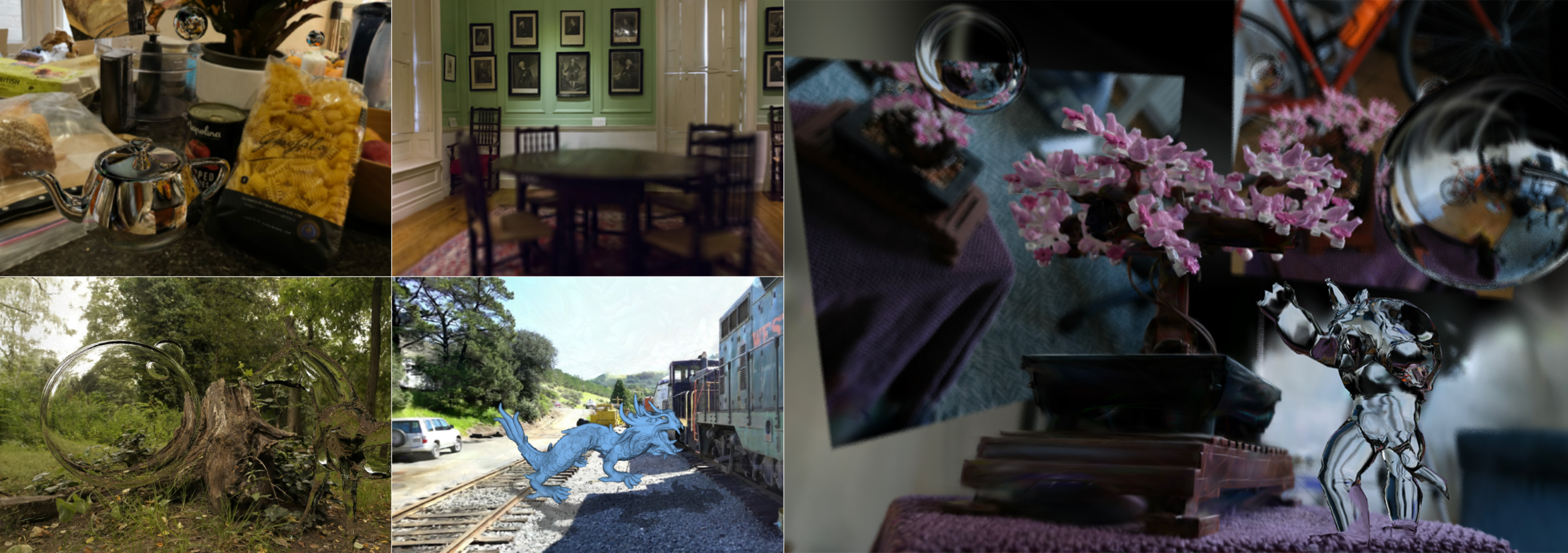}
    \vspace{-6mm}
    \caption{\textbf{Runtime Ray Tracing Effects:} 
    Our ray-based pipeline is easily compatible with conventional ray-based visual effects at test time, including reflections (top left), depth of field (top middle), refractions (bottom left), hard shadows cast by meshes (bottom middle), and myriad combinations of them (right).
    }
    \vspace{-4mm}
    \label{fig:playground_effects}
\end{figure*}

\section{Related Work}

\subsection{Novel-View Synthesis and Neural Radiance Fields}

Classical approaches to novel-view synthesis can be roughly categorized based on the sparsity of the input views. In the case of sparse views, most methods first construct a proxy geometry using multi-view stereo~\cite{schoenberger2016mvs, schoenberger2016sfm} and point cloud reconstruction methods~\cite{ kazdhan2006psr, kazdhan2013spsr} and then unproject the images onto this geometry either directly in terms of RGB colors~\cite{debevec1996hybrid, Buehler2001ulr, wood2000surfacelf} or extracted latent features~\cite{Riegler2020FVS, Riegler2021SVS}. The novel views are rendered by projecting the color or features from the geometry to the camera plane. In the case of densely sampled input views, the novel-view synthesis can instead be formulated directly as light field interpolation problem~\cite{Gortler1996lumigraph, Levoy1996lightfields, davis2012ulf}.

Neural Radiance Fields (NeRFs)~\cite{mildenhall2020nerf} have revolutionized the field of novel-view synthesis by representing the scenes in terms of a volumetric radiance field encoded in a coordinate-based neural network. This network can be queried at any location to return the volumetric density and view-dependent color. The photo-realistic quality of NeRFs has made them the standard representation for novel-view synthesis. Follow-up works have focused on improving the speed~\cite{mueller2022instant, Reiser2021kilonerf}, quality~\cite{barron2021mipnerf, barron2022mipnerf360, barron2023zipnerf}, and surface representation~\cite{wang2021neus, yariv2021volsdf, Li2023Neuralangelo, wang2023fegr}. NeRF has also been extended to large-scale scenes~\cite{Turki2022meganerf, li2024nerfxl}, sparse inputs views~\cite{Niemeyer2021Regnerf}, in-the-wild image collections~\cite{martinbrualla2020nerfw}, \revAdded{and reflections~\cite{guo2022nerfren}}. Finally, several works investigated ways to speed up the inference by baking the neural fields to more performant representations~\cite{reiser2023merf, adaptiveshells2023, duckworth2023smerf, Reiser2024SIGGRAPH}. While achieving high quality and fast rendering speeds, these methods often employ compute-expensive multi-stage training procedures. 

\subsection{Point-Based and Particle Rasterization}
\citet{grossman1998point} defined point-based rendering as a simple rasterization of object surface points along with their color and normals. However, due to the infinitesimal size of the points, such simple rasterization inevitably led to holes and aliasing. To address these limitations, later work converted points to particles with a spatial extent, such as surfels, circular discs, or ellipsoids~~\cite{pfister2000surfels, zwicker2001surface, ren2002object}. More recently, points or particles have also been augmented with neural features and rendered using rasterization in combination with CNN networks~\cite{aliev2020neural, KPLD21, ruckert2022adop} or NeRF-like volumetric rendering~\cite{xu2022point, ost2022neural}. 

Differentiable rendering through alpha blending was also extended to volumetric particles. Pulsar~\cite{Lassner2021PulsarES} proposed an efficient sphere-based differentiable rasterization approach, which allows for real-time optimization of scenes with millions of particles. The seminal 3DGS work of ~\citet{kerbl3Dgaussians} instead represented the scenes using fuzzy, anisotropic 3D Gaussian particles. By optimizing the shape, position, and appearance of these Gaussian particles through an efficient tile-based rasterizer, 3DGS achieves SoTA results in terms of perceptual quality and efficiency. 3DGS inspired many follow-up works that aim to reduce the render time or memory footprint~\cite{niedermayr2023compressed, fan2023lightgaussian, papantonakis2024reducing}, improve surface representation~\cite{guedon2023sugar, Huang2DGS2024}, and support large-scale scenes~\cite{ren2024octree, hierarchicalgaussians24}, and more. 

Jointly, these works have made significant progress, but they still inherit limitations of rasterization. Indeed, they are not able to represent highly distorted cameras, model secondary lighting effects, or simulate sensor properties such as rolling shutter or motion blur. Several works have tried to work around these limitations. ~\citet{niemeyer2024radsplat} first train a Zip-NeRF~\cite{barron2023zipnerf} that can model distorted and rolling shutter cameras and then render perfect pinhole cameras from the neural field and distill them into a 3DGS representation. This allows for fast inference, but is still limited to perfect pinhole cameras. 
\revAdded{To address secondary lighting effects, recent works bake occlusion information into spherical harmonics at each Gaussian~\cite{R3DG2023, liang2023gs} or leverage shading models and environment maps~\cite{jiang2023gaussianshader}.
The latter two of these render only with rasterization; in contrast \citet{R3DG2023} traces rays for initial visibility determination, but uses only a visibilty forward pass, and restricts ray tracing to the training phase, relying on rasterization during inference and inheriting its limitations. 
In contrast, our method uses optimized ray-tracing as the sole renderer throughout both training and inference, which allows for inserting objects, refraction, lens distortion, and other complex effects. 
Additionally, \citet{R3DG2023} use axis-aligned bounding boxes (AABBs) to enclose particles, which results in approximately $3\times$ lower FPS during inference compared to the stretched icosahedrons employed in our optimized tracer (\secref{ablations}).
Finally, for complex lens effects,} \citet{seiskari2024gaussian} model the motion blur and rolling shutter of the camera by approximating them in screen space through rasterization with pixel velocities. Unlike these works, we formulate a principled method for efficient ray tracing of volumetric particles, which natively alleviates all the limitations mentioned above and further allows us to simulate effects such as depth of field and perfect mirrors.  

\subsection{Differentiable Ray Tracing of Volumetric Particles}\label{sec:sotaRayTraceParticles}

Ray tracing became the gold standard for offline rendering of high-quality photo-realistic images~\cite{Whitted1979AnII}. 
The advent of dedicated hardware to efficiently compute the intersection of  camera rays with the scene geometry has also enabled its use for real-time rendering applications such as gaming and the simulation industry. Modern GPUs are exposing ray tracing rendering pipelines, from the computation of dedicated acceleration structures to a specific programmable interface \cite{10.1145/1276377.1276466}.

However, these works are highly optimized for rendering opaque surfaces and efficiently handling order independent semi-transparent surfaces or particles remains challenging \cite{TankiZhang2021}. 

A first class of works \cite{woit2022,mboit2018} proposes to first estimate the transmittance function along the ray and subsequently to integrate the particles' radiance based on this estimate. It assumes the traversal of the full scene to be fast enough; an assumption that does not hold in Gaussian particles for scene reconstruction.

A second class of works consists in re-ordering the particles along the ray. \citet{knoll2019} propose a slab-tracing method to trace semi-transparent volumetric RBF (radial basis function) particles, which enables
real-time ray tracing of scenes consisting of several millions of such particles. 
However, its efficiency is largely based on the assumption of the isotropic shape of the particles and a high level of uniformity in their spatial distribution. 
In \cite{mlat2020}, the multi-layer alpha blending approach from \cite{Salvi2014MultilayerAB} is extended to ray tracing. Their multi-layer alpha tracing supports efficient rendering of any semi-transparent surface but its approximation of the particle's ordering may produce rendering artifacts.

Our formulation takes root in these precursory works. However as opposed to \cite{knoll2019}, it is guaranteed to process every particle intersecting the ray, and contrary to \cite{mlat2020} the hit processing order is consistent, which ensures the differentiability of our tracing algorithm.

Compared to rasterization, differentiable ray tracing of semi-transparent particles has seen much less progress in recent years. Perhaps the most similar rendering formulation to ours was proposed in Fuzzy Metaballs~\cite{keselman2022fuzzy, keselman2023flexible}, but it is limited to scenes with a small set of 3D Gaussian particles (several tens) and images with very low resolution. Different to Fuzzy Metaballs, our method can easily handle scenes with several millions of particles from which it can render full HD images in real-time.
\revAdded{In another direction, \citet{belcour20135d} incorporate defocus and motion blur in to ray tracers by leveraging Gaussian approximations as a sampling technique, rather than a scene representation as used here.}

\section{Background}

We begin with a short review of 3D Gaussian scene representation, volumetric particle rendering, and hardware-accelerated ray tracing.

\subsection{3D Gaussian Parameterization}\label{sec:ParticleBackground}
Extending \citet{kerbl3Dgaussians}, our scenes can be represented as a set of differentiable semi-transparent particles defined by their kernel function.
For example, the kernel function of a 3D Gaussian particle $\ParticleResponse:\mathbb{R}^3\rightarrow\mathbb{R}$ at a given point $\bm{x} \in \mathbb{R}^3$ can be expressed as
\begin{equation}
    \label{eq:gaussian_kernel}
    \ParticleResponse(\bm{x}) = e^{-(\bm{x}-\ParticleCenter)^T\bm{\Sigma^{-1}}(\bm{x}-\ParticleCenter)},
\end{equation}
where $\ParticleCenter \in \mathbb{R}^3$ represents the particle's position and $\bm{\Sigma} \in \mathbb{R}^{3 \times 3}$ the covariance matrix.
To ensure the positive semi-definiteness of the covariance matrix $\bm{\Sigma}$ when optimizing it using gradient descent, we represent it as
\begin{equation}
    \bm{\Sigma}=\bm{R}\bm{S}\bm{S}^T\bm{R}^T
\end{equation}
with $\bm{R} \in \mathrm{SO(3)}$ a rotation matrix and $\bm{S} \in \mathbb{R}^{3 \times 3}$ a scaling matrix. These are both stored as their equivalent vector representations, a quaternion $\ParticleRotation \in \mathbb{R}^4$ for the rotation and a vector $\ParticleScale \in \mathbb{R}^3$ for the scale.
For other particle variants explored in this work, please refer to \secref{generalized_particles}.

Each particle is further associated with an opacity coefficient $\ParticleOpacity \in \mathbb{R}$, and a parametric radiance function $\ParticleRadiance_{\ParticleHarmonics}(\RayDirection):\mathbb{R}^3\rightarrow\mathbb{R}^3$, dependent on the view direction $\RayDirection \in \mathbb{R}^3$.
In practice, following \citet{kerbl3Dgaussians}, we use a spherical harmonics functions $Y_{\ell}^m(\RayDirection)$ of order $m=3$ defined by the coefficients $\ParticleHarmonics \in \mathbb{R}^{48}$. Note that we are using the ray direction while \citet{kerbl3Dgaussians} uses $\frac{\ParticleCenter-\RayOrigin}{\lVert\ParticleCenter-\RayOrigin\rVert}$ for performance reason.

Therefore the parametric radiance function can be written as
\begin{equation}
    \ParticleRadiance_{\ParticleHarmonics}(\RayDirection) = f \left(\sum_{\ell=0}^{\ell_{\max}}\sum_{m=-\ell}^{\ell} \beta_{\ell}^m Y_{\ell}^m(\RayDirection) \right)
    \label{eq:sph}
\end{equation}
where $f$ is the sigmoid function to normalize the colors. %

\subsection{Differentiable Rendering of Particle Representations}
Given this parametrization, the scene can be rendered along a ray $\bm{r}(\tau) = \RayOrigin + \tau\RayDirection$ with origin $\RayOrigin \in \mathbb{R}^3$ and direction $\RayDirection \in \mathbb{R}^3$ via classical volume rendering
\begin{equation}
  \label{eq:VolumeRendering}
    \resizebox{.9\hsize}{!}{$
        \Radiance(\RayOrigin,\RayDirection)=\int_{\tau_n}^{\tau_f}T(\RayOrigin,\RayDirection)\left(\sum_i{(1-e^{-\sigma_i\ParticleResponse_i(\RayOrigin+\tau\RayDirection)})\vc_i(\RayDirection)}\right)d\tau
    $},
\end{equation}
where $\vc_i(\RayDirection)=\ParticleRadiance_{\ParticleHarmonics_i}(\RayDirection)$ is the color of the $i^\text{th}$ Gaussian obtained by evaluating its view-dependant radiance function. The transmittance function $T(\RayOrigin,\RayDirection)$ is defined as 
\begin{equation}
    T(\RayOrigin,\RayDirection) = e^{-\int_{\tau_n}^\tau\sum_i\sigma_i\ParticleResponse_i(\RayOrigin+t\RayDirection)dt}.
\end{equation}
Considering $\alpha_i =\sigma_i\ParticleResponse_i(\mathbf{x}_i)$, \eqref{VolumeRendering} can be approximated using numerical integration as
\begin{equation}
\label{eq:QuadratureVolumeRendering}
\Radiance(\RayOrigin,\RayDirection) = \sum_{i=1}^N \vc_i(\RayDirection) \alpha_i \prod_{j=1}^{i-1} 1 -\alpha_j,
\end{equation}
where in this linear approximation, $\mathbf{x}_i$ is defined as the point along the ray $\bm{r}$ with the highest response $\ParticleResponse_i$ of the $i^\text{th}$ Gaussian (see \ref{sec:ParticleResponseEvaluation} for more details). For derivations of the higher order approximations of $\ParticleResponse_i$ please refer to~\cite{keselman2022fuzzy}.

\begin{figure*}
    \centering
    \includegraphics[width=1.\linewidth]{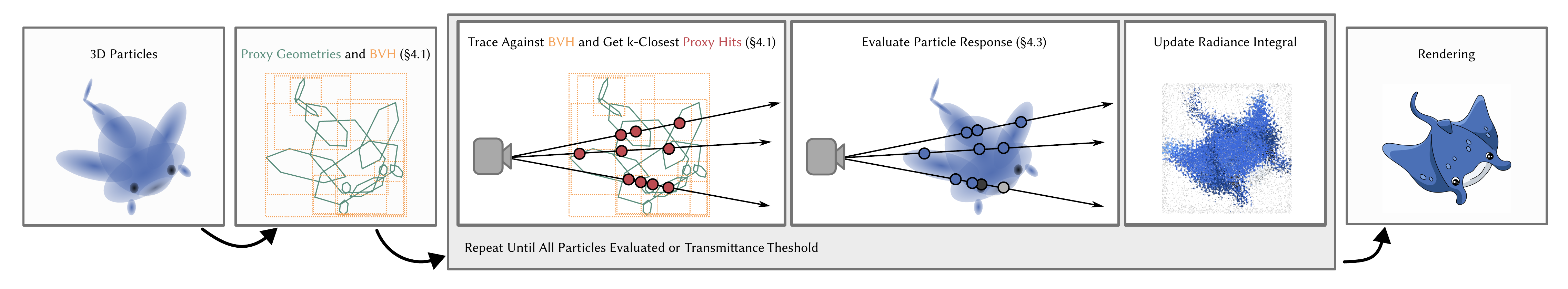}
    \vspace{-9mm}
    \caption{\textbf{Overview of the Accelerated Tracing Algorithm:} Given a set of 3D particles, we first build the corresponding bounding primitives and insert them into a BVH. To compute the incoming radiance along each ray, we trace rays against the BVH to get the next \textit{k} particles. We then compute the intersected particles' response and accumulate the radiance according to~\eqref{QuadratureVolumeRendering}. The process repeats until all particles have been evaluated or the transmittance meets a predefined threshold and the final rendering is returned. 
    }
    \vspace{-3mm}
    \label{fig:method_overview}
\end{figure*}

\begin{figure}
    \centering
    \includegraphics{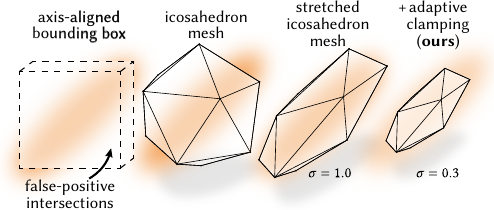}
    \vspace{-3mm}
    \caption{\textbf{Proxy Geometries:} Examples of BVH primitives considered.}
    \label{fig:proxy_geom_diagram}
\end{figure}

\subsection{Hardware-Accelerated Ray Tracing}
\label{sec:HardwareAcceleratedRayTracing}

In this work we use NVIDIA RTX hardware through the NVIDIA OptiX programming interface~\cite{parker2010optix}.
Through this interface, geometric primitives such as triangle meshes are processed to construct a \emph{Bounding Volume Hierarchy} (BVH\footnote{For clarity, throughout this work we reference BVH as the de-facto hardware acceleration structure. However, since in practice NVIDIA OptiX's specification interfaces the implementation of bottom level acceleration structures, we emphasize our pipeline does not depend on a particular implementation.}). 
This acceleration structure is optimized for the computation of ray-primitive intersections by dedicated hardware, the RT cores. 
The programmable pipeline sends traversal queries to this hardware, freeing the GPU streaming-multiprocessors (SMs) for the main computational load, e.g material shading.
The interactions of the SMs with the RT cores are done through the following programmable entry points: 
\begin{itemize}[topsep=4pt, leftmargin=20pt]
    \item \emph{ray-gen} program (ray generation) is where the SMs may initiate a scene traversal for a given ray.
    \item \emph{intersection} program is called during the traversal to compute the precise intersection with potentially hit primitives that are not directly supported by the hardware.
    \item \emph{any-hit} program is called during the traversal for every hit and may further process or reject the hit.
    \item \emph{closest-hit} program is called at the end of the traversal, for further processing of the closest accepted hit.
    \item \emph{miss} program is called at the end of the traversal for further processing when no hit has been accepted.
\end{itemize}
Such a pipeline is highly optimized to render opaque primitives, i.e. the number of expected hits during a traversal is low, with a minimal amount of interactions between the SMs and the RT cores.
Rendering volumes, where the primitives are semi-transparent, requires traversing and processing many hits per ray. To efficiently trace a volume, specific approaches must be designed, tailored to the type of primitives (or particles), their expected size, and distribution across the volume (see for example \cite{knoll2019}). 
In this work we propose an efficient and differentiable algorithm to ray trace a volume made of optimized semi-transparent particles for high-fidelity novel view rendering.

\section{Method}

The proposed volumetric particle tracer requires two core components: a strategy to represent particles in an acceleration structure (BVH) to efficiently test for intersections against them, using adaptive bounding mesh primitives (\secref{BoundingPrimitives}), and a rendering algorithm which casts rays and gathers batches of intersections, efficiently scheduled onto the NVIDIA OptiX tracing model (\secref{RayTracingRenderer}).

\subsection{Bounding Primitives} 
\label{sec:BoundingPrimitives}

Any ray-tracer must somehow insert the primitive particles into a BVH and query the primitives intersected by a ray. The first challenge is then to decide how to insert particles into a BVH and conservatively test intersection against them.

The NVIDIA OptiX programming model supports three primitive types which can be inserted into the BVH: triangles, spheres, and custom primitives given by their axis-aligned bounding boxes (AABBs).
These options admit many possible strategies to build a BVH over particles, such as constructing naive axis-aligned bounds as AABBs or spheres, or building bounding triangle meshes.
These strategies have a tradeoff between the cost to test intersection \vs{} the tightness of the bound.
For instance, simply intersecting a ray with the AABB around each particle is fast, but a diagonally-stretched Gaussian particles will cause the traversal to have to evaluate many false-positive intersections which actually contribute almost nothing to the rendering.
None of these strategies necessarily affect the appearance of the rendered image, but rather the computation speed and number of low-contribution particles needlessly processed.
Billboard mesh proxies are used elsewhere~\cite{niedermayr2023compressed}, but do not apply in our general setting where rays may come from any direction.

\paragraph{Stretched Polyhedron Proxy Geometry}
\label{sec:primitive_clamping}

After experimenting with many variants (\secref{PrimitiveExp}), we find it most effective to encapsulate particles in a stretched regular icosahedron mesh (\figref{proxy_geom_diagram}), which tightly bounds the particle and benefits from hardware-optimized ray-triangle intersections.
A hit against any front-facing triangle of the bounding mesh triggers processing of the corresponding particle, as described later in~\secref{ParticleResponseEvaluation}.
We fit the bounding proxy by specifying a minimum response $\MinResponse$ which must be captured (typically $\MinResponse = 0.01$), and analytically compute an anisotropic rescaling of the icosahedron to cover the whole space with at least $\MinResponse$ response.
Precisely, for each particle we construct an icosahedron with a unit inner-sphere, and transform each canonical vertex $\bm{v}$ according to:
\begin{equation}
  \label{eq:primitive_transform}
  \bm{v} \gets 
  \sqrt{2 \log(\ParticleOpacity / \MinResponse)}
  \bm{S} \bm{R}^{T} \bm{v} + \ParticleCenter.
\end{equation}
Importantly, this scaling incorporates the opacity of the particles, so that large nearly-transparent particles may have smaller bounding primitives, resulting in adaptive clamping of the particles.

\subsection{Ray Tracing Renderer}
\label{sec:RayTracingRenderer}

\paragraph{Motivation}
Given the ability to cast rays against particles, volumetric rendering as in ~\eqref{QuadratureVolumeRendering} requires accumulating the contribution of particles along the ray in a sorted order.
One naive approach within the NVIDIA OptiX programming model (\secref{HardwareAcceleratedRayTracing}) is to repeatedly cast the ray, process the nearest particle with a \emph{closest-hit} program, then re-cast the ray to find the next particle.
Another is to traverse the scene only twice, once to estimate the transmittance function, and once to the compute the integral as in \cite{mboit2018}. 
Both of these strategies are prohibitively expensive, due to the cost of traversing the scene.

Our renderer builds on past approaches for tracing semi-transparent surfaces or particles: ~\citet{knoll2019} repeated gather slabs of particles and sort within each slab, while ~\citet{mlat2020} process all semi-transparent surfaces into a $\ParticleTraceBatchSize$-buffer, merging adjacent particles when the list overflows.
As discussed in \secref{sotaRayTraceParticles}, because of their approximations, these algorithms do not produce a consistent rendering, which prevents differentiation and generates artifacts.

\begin{figure}
    \centering
    \includegraphics[width=.95\linewidth]{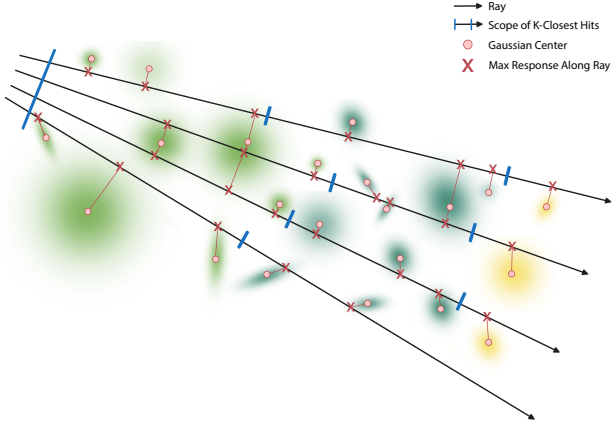}
    \vspace{-4mm}
    \caption{\textbf{Rendering:} on each round of tracing, the next $\ParticleTraceBatchSize$ closest hit particles are collected and sorted into depth order along the ray, the radiance is computed in-order, and the ray is cast again to \revAdded{collect the next hits}.
    }
    \label{fig:slabs}
\end{figure}

\paragraph{Algorithm}

\OPC{
\figref{slabs},
} \figref{method_overview}, \procref{raygen}, and \procref{anyhit} summarize our approach. 
To compute incoming radiance along each ray, a \emph{ray-gen} program traces a ray against the BVH to gather the next $\ParticleTraceBatchSize$ particles, using an \emph{any-hit} program to maintain a sorted buffer of their indices.
For efficiency, at this stage the particle response is not yet evaluated; all primitive hits are treated as intersected particles.
The \emph{ray-gen} program then iterates through the sorted array of primitive hits, retrieves the corresponding particle for each, and renders them according to~\eqref{QuadratureVolumeRendering}.
The process then repeats, tracing a new ray from the last rendered particle to gather the next $\ParticleTraceBatchSize$ particles.
The process terminates once all particles intersecting the ray are processed, or early-terminates as soon as enough particle density is intersected to reach a predefined minimum transmittance $\MinTransmittance$.
Compared to past approaches this renderer allows for processing the intersection in a consistent order, without missing any particle nor approximating the transmittance.

Nonetheless, this proposed algorithm is just one of many possible variants, chosen for performance after extensive benchmarking.
See ~\secref{ablations} for timings and ablations against a selection of alternatives considered; we find that subtle changes to the algorithm have a dramatic effect of speed and quality on densely-clustered multi-view scenes.

\setlength{\columnsep}{1em}
\setlength{\intextsep}{0em}
\begin{wrapfigure}{r}{100pt}
  \includegraphics{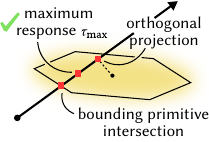}
\end{wrapfigure}

\subsection{Evaluating Particle Response}
\label{sec:ParticleResponseEvaluation}
After identifying ray-particle intersections, we must choose how to compute the contribution of each particle to the ray.
As with prior work, we take a single sample per particle, but we still must choose at what distance $\tau$ along the ray to evaluate that sample.
\citet{knoll2019} orthogonally project the center of the particle on to the ray; this strategy is reasonable for isotropic particles, but can lead to significant error for stretched anisotropic particles.
Instead, we analytically compute a sample location $\TMax=\textrm{argmax}_{\tau}{\ParticleResponse(\RayOrigin+\tau \RayDirection)}$, the point of maximum response from the particle along the ray.
For Gaussian particles, this becomes
\begin{equation}
  \label{eq:GaussianMaxResponse}
  \TMax = \frac{(\ParticleCenter - \RayOrigin)^T\bm{\Sigma}^{-1}\RayDirection}{\RayDirection^T\bm{\Sigma}^{-1}\RayDirection} = \frac{-\RayOrigin_g^T\RayDirection_g}{\RayDirection_g^T \RayDirection_g}
\end{equation}
where $\RayOrigin_g=\bm{S}^{-1}\bm{R}^T(\RayOrigin-\ParticleCenter)$ and $\RayDirection_g=\bm{S}^{-1}\bm{R}^T\RayDirection$.

Note that this strategy incurs a slight approximation in the ordering: the particle hits are integrated in the order of the bounding primitive intersections instead of the order of the sample locations. However, we empirically confirmed that this approximation does not lead to any substantial loss in the quality of the end result.

\subsection{Differentiable Ray Tracing and Optimization}

\paragraph{Differentiable Rendering}
Beyond forward-rendering of particle scenes, our ray tracing renderer is also \emph{differentiable}, to support optimizing particle scenes from observed data.
To backpropagate (\ie{}, reverse-mode differentiate) through the renderer with respect to particle parameters, we first perform an ordinary forward-pass render and compute the desired objective functions.
Then, in the backward pass we re-cast the same rays to sample the same set of particles in order, computing gradients with respect to each shading expression and accumulating gradients in shared buffers with atomic scatter-add operations.
We implemented all derivative expressions by hand in an NVIDIA OptiX \emph{ray-gen} program which is structurally similar to ~\procref{raygen}.

\noindent
\begin{minipage}{\columnwidth} %
\vspace{2em}
  \begin{algo}{\Proc{Ray-Gen}$(\RayOrigin, \RayDirection, \MinTransmittance, \MinAlphaResponse, \ParticleTraceBatchSize, \SceneTMin, \SceneTMax)$}
  \label{proc:raygen}
  \begin{algorithmic}[1]
    \InputConditions{
ray origin $\RayOrigin$, ray direction $\RayDirection$, min transmittance $\MinTransmittance$, min particle opacity $\MinAlphaResponse$, hit buffer size $\ParticleTraceBatchSize$, ray scene bounds $\SceneTMin$ and $\SceneTMax$}
    \OutputConditions{ray incoming radiance $\Radiance$, ray transmittance $\Transmittance$}
    
    \State $\Radiance \gets (0.,0.,0.) $ \Comment{radiance}
    \State $\Transmittance \gets 1. $ \Comment{transmittance}
    \State $\CurrT \gets \SceneTMin $ \Comment{Minimum distance along the ray}
    \While{$\CurrT < \SceneTMax$ and $\Transmittance > \MinTransmittance$}
      
      \LeftComment{Cast a ray to the BVH for the next $\ParticleTraceBatchSize$ hits, sorted}
      \State $\mathcal{H} \gets \text{TraceRay}(\RayOrigin + \CurrT \RayDirection,\RayDirection, \ParticleTraceBatchSize)$\\

      \For{$(\THit, i_\textrm{prim}) \text{ in } \mathcal{H}$}
      \Comment{Render this batch of hits}

        \State $i_\textrm{particle} \gets \text{GetParticleIndex}(i_\textrm{prim})$
        \State $\AlphaResponse_\textrm{hit} \gets \text{ComputeResponse}(\RayOrigin, \RayDirection, i_\textrm{particle})$ \Comment{$\ParticleOpacity \ParticleResponse(\RayOrigin + \tau \RayDirection)$}

        \If{$\AlphaResponse_\textrm{hit} > \MinAlphaResponse$}

          \State $\Radiance_\textrm{hit} \gets \text{ComputeRadiance}(\RayOrigin, \RayDirection, i_\textrm{particle})$ 
          \Comment{Refer to \eqref{sph} for SH evaluation}

          \State $\Radiance \gets \Radiance + \Transmittance * \AlphaResponse_\textrm{hit} * \Radiance_\textrm{hit}$
          \State $\Transmittance \gets \Transmittance * (1 - \AlphaResponse_\textrm{hit})$
          
        \EndIf

        \State $\CurrT \gets \THit$ \Comment{Resume tracing from last hit}
      \EndFor

    \EndWhile
      
    \State \textbf{return} $\Radiance$, $\Transmittance$
  \end{algorithmic}
\end{algo}
\end{minipage}

\noindent
\begin{minipage}{\columnwidth} %
\vspace{2em}
  \begin{algo}{\Proc{Any-Hit}$(\THit, i_\textrm{prim}, \mathcal{H}, \ParticleTraceBatchSize)$}
  \label{proc:anyhit}
  \begin{algorithmic}[1]
    \InputConditions{hit location $\THit$, primitive index $i_\textrm{prim}$, hit array $\mathcal{H}$, hit buffer size $\ParticleTraceBatchSize$}
    \OutputConditions{the hit array $\mathcal{H}$ may be updated in-place with a new entry}
   
    \State h $\gets (\THit, i_\textrm{prim})$

    \For{$i \text{ in } 0 \dots \ParticleTraceBatchSize$-1} \Comment{insertion sort into hit array}

      \If{$ h.\THit < \mathcal{H}[i].\THit $}
        \State $\text{swap}(\mathcal{H}[i], h)$
      \EndIf
    \EndFor
    \\
    \LeftComment{ignore $\ParticleTraceBatchSize$-closest hits to prevent the traversal from stopping}
    \If{$\THit < \mathcal{H}[\ParticleTraceBatchSize-1].\THit$}
      \State $\text{IgnoreHit()}$
    \EndIf
  \end{algorithmic}
\end{algo}
\end{minipage}

\paragraph{Optimization}\label{par:optimization}
To fit particle scenes using our ray tracer, we adopt the optimization scheme of \citet{kerbl3Dgaussians}, including pruning, cloning and splitting operations. 
One significant change is needed: \citet{kerbl3Dgaussians} track screen-space gradients of particle positions as a criteria for cloning and splitting, but in our more-general setting, screen space gradients are neither available nor meaningful---instead, we use gradients in 3D world-space for the same purpose.
Recent work has proposed many promising extensions to the optimization scheme of \citet{kerbl3Dgaussians}. While our ray tracer is generally compatible with any of these extensions, we stay faithful to the original approach for the sake of consistent comparisons.
It should also be noted that as particles are updated during optimization, the ray tracing BVH must be regularly rebuilt (see \figref{inferenceAlgoComparison}, bottom left for BVH build time).

\paragraph{Training with Incoherent Rays}\label{par:incoherentrays}
Optimization in computer vision often benefits from stochastic descent, fitting to randomly-sampled subsets of a problem on each iteration.
However, differentiable rasterization can only efficiently render whole images or tiles, and thus efficient stochastic optimization over the set of pixels in a scene is not possible.
In our ray tracer, we are free to train with \emph{stochastically-sampled} rays, drawn at random or according to some importance sampling during training, see ~\secref{benchmarks}.
Note that when stochastic sampling is used, window-based image objectives such as SSIM cannot be used.

\subsection{Particle Kernel Functions}
\label{sec:generalized_particles}
\begin{figure}
    \centering
    \includegraphics[width=\linewidth]{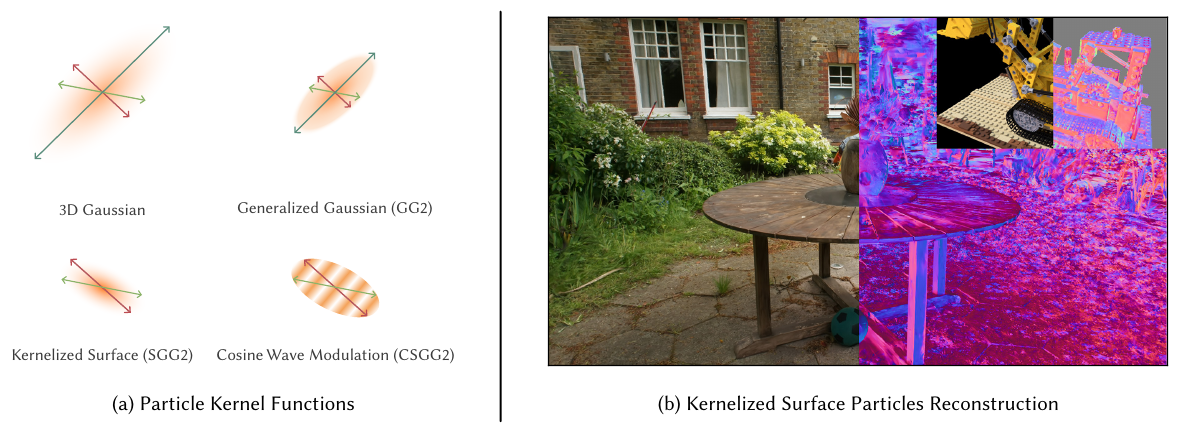}
    \vspace{-6mm}
    \caption{\textbf{Particle Kernel Functions:} (a) In addition to 3D Gaussians, in this work we investigated three other particle types: the generalized Gaussian (GG2), kernelized surface (SGG2) and cosine wave modulation (CSGG2) particles. (b) Shows radiance and normal reconstructions obtained with the kernelized surface particles for two scenes.}
    \label{fig:particle_kernel_functions}
\end{figure}

Our formulation does not require the particles to have a Gaussian kernel, enabling the exploration of other particle variants.
We consider a general particle defined by its kernel function $\hat{\ParticleResponse}(\bm{x})$.
In addition to the standard Gaussian, we investigate three other variants, visualized in ~\figref{particle_kernel_functions}:
\begin{itemize}[topsep=4pt, leftmargin=20pt]
  \item The standard 3D Gaussian kernel given in \eqref{gaussian_kernel} as
\begin{equation*}
    \hat{\ParticleResponse}(\bm{x}) = \ParticleOpacity e^{-(\bm{x}-\ParticleCenter)^T\bm{\Sigma^{-1}}(\bm{x}-\ParticleCenter)},
\end{equation*}
    \item Generalized Gaussians of degree $n$ (we use $n=2$): 
    \begin{equation}
        \hat{\ParticleResponse}_n(\bm{x}) = \ParticleOpacity e^{-((\bm{x}-\ParticleCenter)^T\bm{\Sigma^{-1}}(\bm{x}-\ParticleCenter))^n}  .  
    \end{equation}
  \item Kernelized surfaces: 3D Gaussians with a null $\bm{z}$ scale as in~\cite{Huang2DGS2024}.
  \item Cosine wave modulations along axis $i$: 
    \begin{equation}
        \hat{\ParticleResponse}_{c}(\bm{x}) =\hat{\ParticleResponse}(\bm{x})(0.5 + 0.5cos(\ParticleCosineModulation (\bm{S}^{-1}\bm{R}^T(\bm{x}-\ParticleCenter))_i))
    \end{equation}
    with $\ParticleCosineModulation$ an optimizable parameter.
\end{itemize}
Comparative evaluations with these particles are presented in \tabref{generalized_particles}.
The generalized Gaussian kernel function defines denser particles, reducing the number of intersections and increasing the performance by a factor of 2 compared to standard Gaussians, see \secref{KernelFuncExp} for more discussion.
The kernelized surface variant defines flat particles with well-defined normals, which can be encapsulated by a two triangle primitive (\secref{BoundingPrimitives}) well-adapted to our tracing model.
Finally, the modulation by a cosine wave \OPC{aims to model} a particle with spatially varying radiance.

\begin{table*}[]
\caption{Results for our approach and baselines on a variety of novel view synthesis benchmarks.}
\footnotesize
\vspace{-3mm}
\begin{center}
\begin{tabular}{l|llll|llll|llll}
\toprule
                      & \multicolumn{4}{c|}{\texttt{MipNeRF360}}                                                     & \multicolumn{4}{c|}{\texttt{Tanks \& Temples}}                                                  & \multicolumn{4}{c}{\texttt{Deep Blending}} \\
Method\textbackslash{}Metric & \multicolumn{1}{c}{PSNR$\uparrow$} & \multicolumn{1}{c}{SSIM$\uparrow$} & \multicolumn{1}{c}{LPIPS$\downarrow$} &\multicolumn{1}{c|}{Mem. $\downarrow$} & \multicolumn{1}{c}{PSNR$\uparrow$} & \multicolumn{1}{c}{SSIM$\uparrow$} & \multicolumn{1}{c}{LPIPS$\downarrow$} & \multicolumn{1}{c|}{Mem. $\uparrow$} & \multicolumn{1}{c}{PSNR$\uparrow$} & \multicolumn{1}{c}{SSIM$\uparrow$} & \multicolumn{1}{c}{LPIPS$\downarrow$} &\multicolumn{1}{c}{Mem. $\downarrow$} \\ \midrule
Plenoxels                    &  23.63  &0.670  & -  & 2.1GB &   21.08  & 0.719 & - &  2.3GB & 23.06 & 0.795 & - & 2.7GB\\

INGP-Base                    &  26.43  &0.725  & -  & 13MB &   21.72  & 0.723 & - & 13MB & 23.62 & 0.797 & - & 13MB\\   
INGP-Big                     &  26.75  &0.751  & -  &  48MB &   21.92  & 0.745 & - & 48MB & 24.96 & 0.817 & - & 48MB\\

MipNeRF360                   &  29.23  & 0.844 & - &  8.6MB &   22.22  & 0.759 & - &  8.6MB & 29.40 & 0.901 & - &8.6MB\\
\revAdded{Zip-NeRF}                   &  \revAdded{30.38} & \revAdded{0.883} & 0.197 & - & - & - & - & -  & - & - & - & -\\
\midrule
3DGS (paper)                 &  28.69  & 0.871 & - & 734MB  &   23.14  & 0.853 & - & 411MB & 29.41 & 0.903 & - & 676MB\\
3DGS (checkpoint)            &  28.83  &0.867  & 0.224 &  763MB &   23.35  & 0.837 & - &   422MB & 29.43 & 0.898 & - &  704MB\\
\midrule
Ours  (reference)          &  28.69  &0.871  & 0.220 & 387MB &  23.03 & 0.853 & 0.193 & 519MB &29.89 & 0.908 & 0.303 & 329MB\\

Ours &  28.71  &0.854  & 0.248  & 383MB &   23.20 & 0.830 & 0.222 & 489MB & 29.23 & 0.900 & 0.315 & 287MB\\
\bottomrule
\end{tabular}
\end{center}
\label{tab:main_benchmark}
\vspace{-3mm}
\end{table*}

\section{Experiments and Ablations}
In this section we evaluate the proposed approach on several standard benchmark datasets for quality and speed, and perform ablation studies on key design choices in \secref{ablations}.
Additional details on experiments and implementation can be found in the appendix.

\paragraph{Method Variants}
In the experiments that follow, we will refer to two variants of our method.
The \emph{Ours (reference)} variant corresponds to \cite{kerbl3Dgaussians} as closely as possible. It employs regular 3D Gaussian particles, and leaves the optimization hyperparameters unchanged. We treat this as a high-quality configuration.
The \emph{Ours} variant is adapted based on the experiments that follow, improving runtime speed at a slight expense of quality.
It uses degree-2 generalized Gaussian particles, a density learning rate to $0.09$ during optimizing, as well as optimizing with incoherent rays in a batch size of $2^{19}$ starting after 15,000 training iterations. 
\revAdded{Empirically, we find that the larger density learning rate of this model produces denser particles. When combined with the faster fall-off of degree-2 generalized Gaussian particles compared to regular Gaussians, this leads to fewer hits along the rays and faster rendering speeds with minimal quality loss.}

\begin{figure*}[!t]
    \centering
    \includegraphics[width=0.98\textwidth]{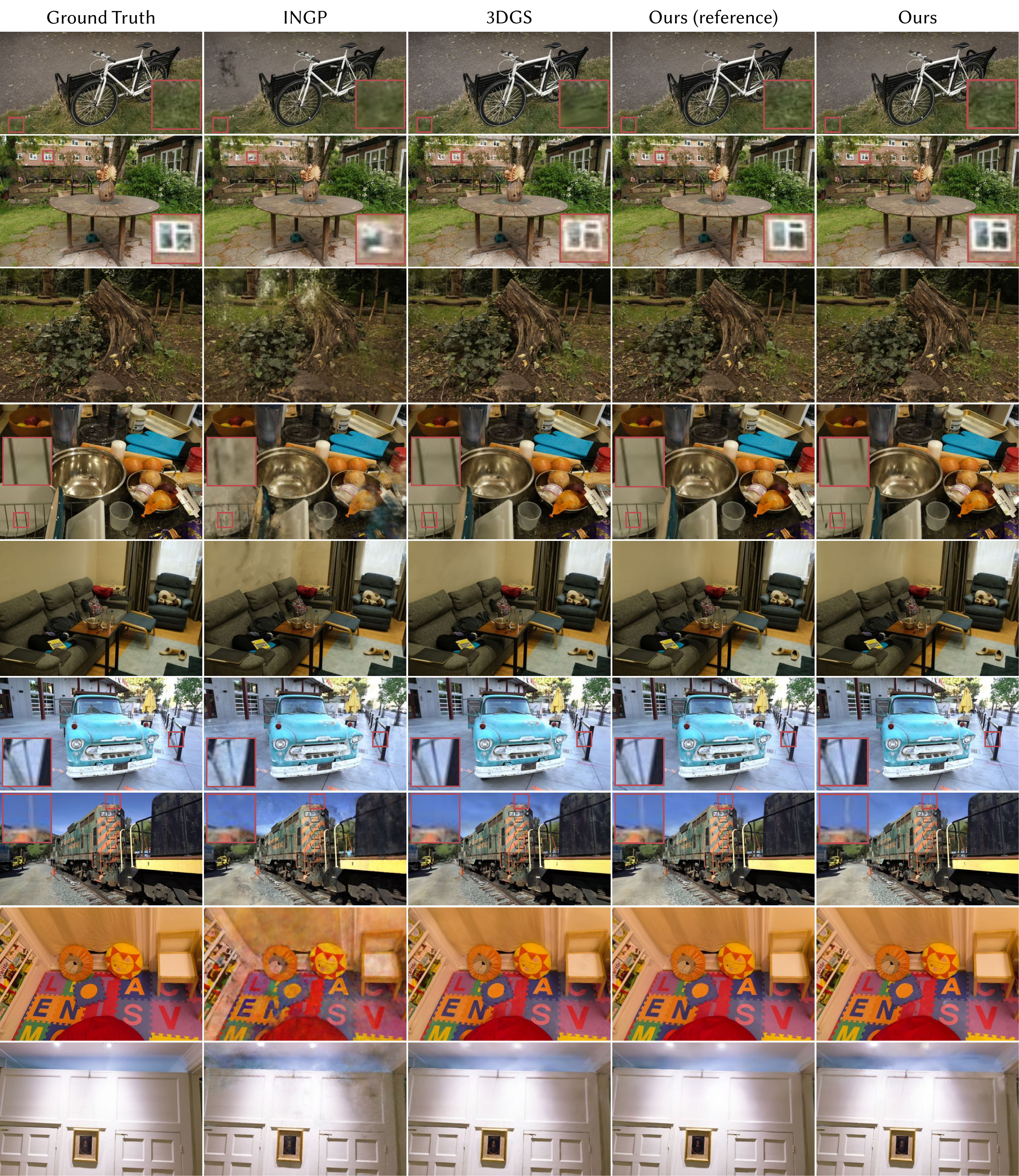}
    \caption{
    \textbf{Novel-View Synthesis:} Qualitative comparison of our
    novel-view synthesis results relative to baselines (insets
    ({\color{palette4} $\bullet$}) show per-result closeups). \OPC{For fairness, this comparison uses the same test views picked by \cite{kerbl3Dgaussians}. Additional comparisons with \cite{barron2022mipnerf360} are included in the appendix.}
    }
    \label{fig:results_gallery}
\end{figure*}

\subsection{Novel View Synthesis Benchmarks}
\label{sec:benchmarks}
\paragraph{Baselines}
There is a significant amount of recent and ongoing research on scene representation. We include comparisons to several representative well-known methods, including 3DGS~\cite{kerbl3Dgaussians}, INGP~\cite{mueller2022instant}, and MipNeRF360~\cite{barron2022mipnerf360}, as a standard for comparison.
The latter two are widely-used ray-marching methods that, like this work, do not have the limitations of rasterization. 
We additionally compare with the non-neural method of Plenoxels~\cite{yu_and_fridovichkeil2021plenoxels}.

\paragraph{Evaluation Metrics}
We evaluate the perceptual quality of the novel-views in terms of peak signal-to-noise ratio (\emph{PSNR}), learned perceptual image patch similarity (\emph{LPIPS}), and structural similarity (\emph{SSIM}) metrics. To evaluate the efficiency, we measure GPU time required for rendering a single image without the overhead of storing the data to a hard drive or visualizing them in a GUI. Specifically, we report the performance numbers in terms of frames-per-second measured on a single NVIDIA RTX 6000 Ada. For all evaluations, we use the dataset-recommended resolution for evaluation.

\subsubsection{MipNeRF360} \texttt{MipNeRF360}~\cite{barron2022mipnerf360} is a challenging dataset consisting of large scale outdoor and indoor scenes. In our evaluation, we use the four indoor (\texttt{room}, \texttt{counter}, \texttt{kitchen}, \texttt{bonsai}) and three outdoor (\texttt{bicycle}, \texttt{garden}, \texttt{stump}) scenes without licensing issues. In line with prior work, we used images downsampled by a factor two for the indoor and a factor four for the outdoor scenes in all our evaluations.

\tabref{main_benchmark} shows quantitative results, while novel-views are qualitatively compared in \figref{results_gallery}. In terms of quality, our method performs on par or slightly better than 3DGS~\cite{kerbl3Dgaussians} and other state-of-the-art methods. 
\OPC{
For this dataset, we also compare our method against the recent top-tier method of Zip-NeRF~\cite{barron2023zipnerf} which achieves 30.38 PSNR. 
In terms of runtime (\tabref{fps_benchmark}), at 78 FPS our efficient ray tracing implementation is approximately three times slower than rasterization (238 FPS), while maintaining interactive speeds compared to high-quality ray-marching based works MipNeRF360 and Zip-NeRF (<1 FPS).
\revAdded{
Zip-NeRF employs multisampling to approximate conical frustums in the ray-casting process, combining MipNeRF360’s anti-aliasing techniques with the speedup mechanism of INGP. While achieving unprecedented rendering quality, Zip-NeRF does not support real-time rendering ($<1$ FPS).
}
}
 
\subsubsection{Tanks \& Temples} \texttt{Tanks \& Temples} dataset contains two large outdoor scenes (\texttt{Truck} and \texttt{Train}) with a prominent object in the center, around which the camera is rotating. These scenes are particularly challenging due to the illumination differences between individual frames as well as the presence of dynamic objects. 

Similar to the results on \texttt{MipNeRF360} dataset, our method again performs on par with the state-of-the-art methods in terms of quality, while the speed is approximately 1.7 times slower than 3DGS at 190 FPS. Qualitative results are depicted in \figref{results_gallery}. On this dataset, \emph{Ours} achieves better PSNR than our \emph{Ours (reference)}, but is still worse in terms of LPIPS and SSIM. We hypothesize that this is due to the lack of SSIM loss supervision when training the model with incoherent rays.

\subsubsection{Deep Blending} Following ~\citet{kerbl3Dgaussians}, we use two indoor scenes \texttt{Playroom} and \texttt{DrJohnson} from the \texttt{Deep Blending} dataset. \tabref{main_benchmark} shows that our reference implementation \emph{Ours (reference)} outperforms all baselines across all qualitative metrics. Different to other datasets, we observe a larger quality drop of \emph{Ours}. This is a result of a quality drop on \texttt{Playroom} where we observed instability of the training with incoherent rays. We leave more detailed investigation and parameter tuning of incoherent rays training for future work.

\subsubsection{NeRF Synthetic} \texttt{NeRF Synthetic} is a commonly used synthetic object-centring dataset introduced by ~\citet{mildenhall2020nerf}. The eight scenes with synthetic objects were rendered in Blender and display strong view-dependent effects and intricate geometry structures. See \tabref{nerf_synthetic} in the appendix for a per-scene breakdown.

Both versions of our method outperform all the baselines in terms of PSNR. In fact, \emph{Ours} outperforms \emph{Ours (reference)} on seven out of eight scenes. We conjecture this is due to the simplicity of scenes which are well represented with less hits, and the positive contribution of training with incoherent rays. On these simpler scenes with lower resolution images, our method is capable of rendering novel views at 450FPS and is only 50\% slower than 3DGS.

\begin{table}[]
\caption{Rendering performance: rasterization v.s. ray tracing.}
\vspace{-3mm}
\footnotesize
\begin{center}
\begin{tabular}{l|l|l|l}
\toprule

& \multicolumn{3}{c}{FPS$\uparrow$}  \\

Method & \multicolumn{1}{c|}{\texttt{MipNeRF360}} & \multicolumn{1}{c|}{\texttt{Tanks \& Temples}} & \multicolumn{1}{c}{\texttt{Deep Blending}} \\

\midrule

3DGS (checkpoint) & 238 & 319 & 267 \\

\midrule
Ours (reference) & 55 & 143 & 77 \\
Ours & 78 & 190 & 119 \\

\bottomrule
\end{tabular}
\end{center}
\label{tab:fps_benchmark}
\vspace{-5mm}
\end{table}

\revAdded{
\subsubsection{Zip-NeRF and Distortion}
The rasterization approach in 3DGS \cite{kerbl3Dgaussians} implicitly treats supervision images as coming from perfect-pinhole cameras, whereas our ray-tracing approach can easily be applied directly to highly-distorted cameras such as fisheye captures.
Images can be undistorted through postprocessing to enable fitting with 3DGS, but this comes at a cost, including significant cropping or wasted space in the image plane.

We demonstrate this effect on the Zip-NeRF dataset~\cite{barron2023zipnerf}, which includes four large-scale scenes featuring both indoor and outdoor areas.
These scenes are originally captured from highly distorted fisheye cameras, with undistorted versions also provided through postprocessing. 
\tabref{zipnerf_scenes_comparison} and \figref{zipnerf_scenes_comparison} compare the quality of training 3DGS on undistorted views, \vs{} our ray-traced method on the undistorted views or the original distorted images.
Our model achieves the highest quality when trained on the original distorted images, partly due to the loss of input supervision signals caused by the cropping of marginal pixels during undistortion. 
Note that rendering 3DGS from the original fisheye views is impossible, as its tile-based rasterizer is designed for perfect-pinhole rendering.

\begin{figure*}[!t]
    \centering
    \includegraphics[width=0.98\textwidth]{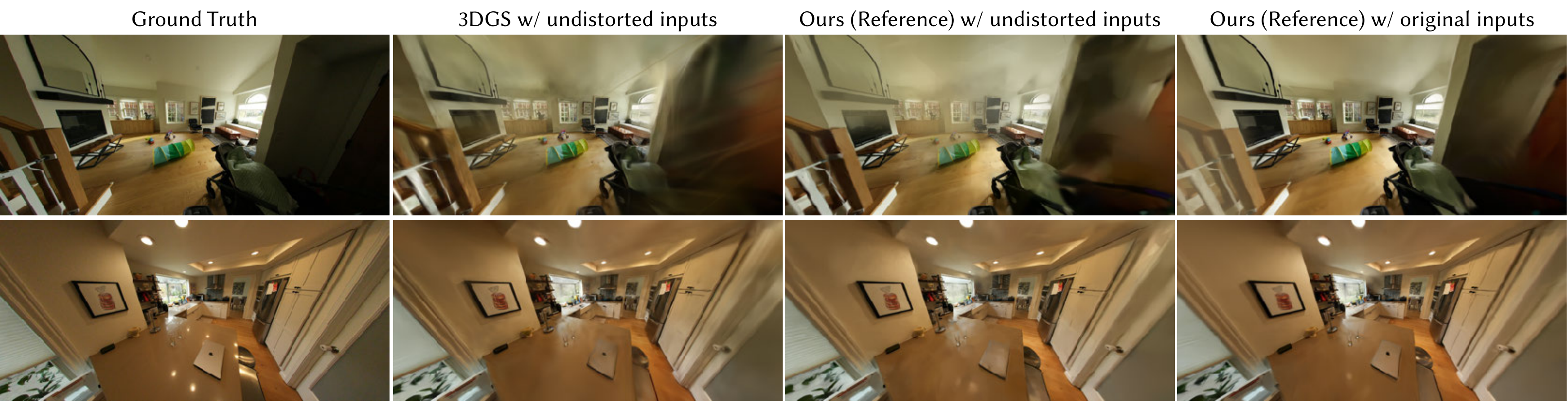}
    \caption{\revAdded{
    Qualitative comparison of our results against 3DGS~\cite{kerbl3Dgaussians} when trained on distorted or undistorted views and then rendered from undistorted views.
    }}
    \vspace{3mm}
    \label{fig:zipnerf_scenes_comparison}
\end{figure*}

}

\subsection{Ray Tracing Analysis and Ablations}
\label{sec:ablations}

We evaluate the performance of the proposed ray tracer and compare to alternatives.
Experiments are evaluated on the union of all validation datasets from ~\secref{benchmarks}.
We measure forward rendering time, which we observed to correlate closely with the per-iteration time for the backward pass.

\subsubsection{Particle Primitives}
\label{sec:PrimitiveExp}

We first consider different bounding primitive strategies as discussed in~\secref{BoundingPrimitives}.
The primitives evaluated are: 
\begin{itemize}[topsep=4pt, leftmargin=20pt]
    \item \textbf{Custom primitive AABBs}: bounding box primitive, see \figref{proxy_geom_diagram} left.
    \item \textbf{Octahedron}: an eight-faced regular polyhedron mesh.
    \item \textbf{Icosahedron}: a twenty-faced regular polyhedron mesh.
    \item \textbf{Icosahedron + unclamped}: icosahedron without adaptive clamping.
\end{itemize}
Scales are determined as in ~\eqref{primitive_transform}, except the \emph{unclamped} variant which omits the opacity term in that expression.

\begin{table}[]
\caption{\revAdded{
Comparison of PSNR achieved by our method versus 3DGS~\cite{kerbl3Dgaussians} when trained and tested on distorted or undistorted views.
}}
\vspace{-3mm}
\footnotesize
\begin{center}
\resizebox{1.0\columnwidth}{!}{
\begin{tabular}{l|l|l}
\toprule

& \multicolumn{2}{c}{Test views}  \\

Method & \multicolumn{1}{c|}{{undistorted}} & \multicolumn{1}{c}{{original (fisheye)}} \\

\midrule

3DGS w/ undistorted inputs & 24.18 & N/A \\
\midrule
Ours (reference) w/ undistorted inputs & 24.59 & 23.69 \\
Ours (reference) w/ original inputs & 24.71 & 24.40 \\

\bottomrule
\end{tabular}
}
\end{center}
\label{tab:zipnerf_scenes_comparison}
\vspace{-5mm}
\end{table}

\figref{inferenceAlgoComparison} (bottom-left) shows the time to build a BVH relative to the number of particles for the different primitives. For simple AABBs, the build time is almost constant whereas for the more complex icosahedron based primitives, the build time is close to linear with more than 30ms per millions of particles.
The same figure also gives the framerate \vs{} the number of particles for different primitives. 
First, the \emph{number} of particles does not strictly determine the performance.
Second, more complex primitives with tighter bounding envelopes yield higher framerates, and adaptive clamping based on opacity has a large positive effect.

\subsubsection{Tracing Algorithms}
\label{sec:tracing-algo}

We consider several alternatives of the proposed ray tracer from~\secref{RayTracingRenderer}, both comparisons to prior work and variants of our method. 
The evaluated approaches are:
\begin{itemize}[topsep=4pt, leftmargin=20pt]
  \item \textbf{Naive closest-hit tracing}: repeated \emph{closest-hit} ray-tracing to traverse every particle hitting the ray in depth order.
  \item \textbf{Slab tracing \cite{knoll2019} (SLAB)}: tracing \revAdded{slabs along the ray}, order independent collection of the $\ParticleTraceBatchSize$-first hits in the \emph{any-hit} program, sorting and integrating the hits in the \emph{ray-gen} program.
  \item \textbf{Multi-layer alpha tracing \cite{mlat2020} (MLAT)}: tracing a single ray with in depth order collection of the hits and merging the closest hits when the $\ParticleTraceBatchSize$ buffer is full in the \emph{any-hit} program.
  \item \textbf{Our proposed method}: tracing a dynamic number of rays with in-order collection of the hits, stopping to evaluate contributions when the $\ParticleTraceBatchSize$ buffer is full in the \emph{any-hit} program. 
  \item \textbf{Ours + tiled tracing}: tracing one ray per $N\times N$ tile, but still evaluating appearance per pixel, akin to tile-based rasterization.
  \item \textbf{Ours + stochastic depth sampling}: tracing a single ray with in depth order collection of the $\ParticleTraceBatchSize$ first accepted samples. Samples are accepted based on the importance sampling $q(\bm{x})=\hat{\ParticleResponse}(\bm{x})$.
\end{itemize}
For each \revReplaced{algorithms}{algorithm} with a choice of parameters (size of the array, number of slabs, or number of samples), we perform a parameter sweep and present the best-performing setting.

\begin{figure}
    \centering
    \includegraphics[width=\linewidth]{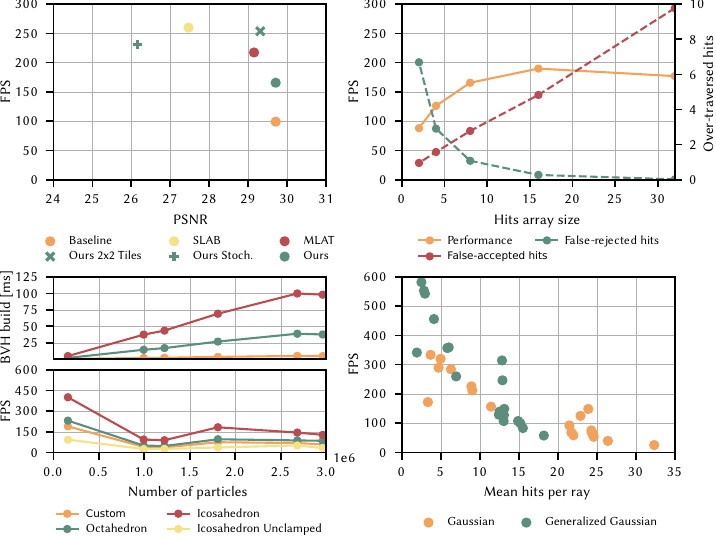}
    \vspace{-6mm}
    \caption{\textbf{Quantitative Ablation.} Top left: comparison of the different tracing algorithms on the combination of our datasets. Top right: Impact of the hit payload buffer size on our proposed tracing algorithm. Bottom left: Impact of the different primitives on both the BVH building time and the FPS. Bottom right: Mean number of hits vs. mean FPS for every sequence of our dataset.}
    \label{fig:inferenceAlgoComparison}
\end{figure}

The performance relative to the accuracy of these implementations are shown in the top-left of \figref{inferenceAlgoComparison}. 
Naive closest-hit tracing is almost twice as slow as our method, due to over-traversal of the BVH.
Slab tracing and multi-layer alpha tracing are designed to minimize over-traversal and hence achieve much better runtime performance, but this speed comes from approximate image formation (ordering approximation for MLAT, under-sampling particles for SLAB), and the accuracy of these methods is substantially lower.
In the differentiable setting, we find that these these approximations make those methods unusable for optimizing scenes.
Adding tile-based rendering to our approach yields a promising speedup, at the cost of a small approximation.
We do not see immediate benefit from stochastic depth sampling, because most of the computation has to be done in the \emph{any-hit} program, preventing a good balance of the tracing workload.

\medskip

\vspace{2mm}

\subsubsection{Particle Kernel Function}
\label{sec:KernelFuncExp}

In \secref{generalized_particles} we consider particle kernel functions beyond Gaussians.
\tabref{generalized_particles} gives results; notably generalized Gaussians with $n=2$ significantly increase ray tracing speed at only a small cost of quality.

\begin{figure}
    \centering
    \includegraphics[width=\linewidth]{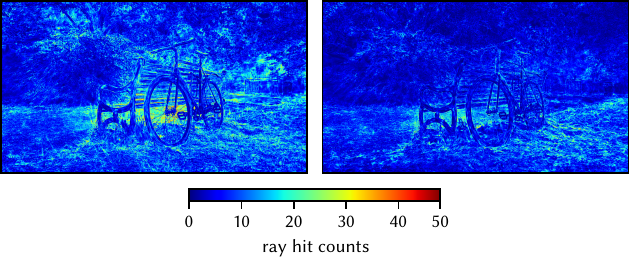}
    \vspace{-6mm}
    \caption{\textbf{Ray Hits for Kernel Functions:} Visualization of the number of per-ray particles
    hits for the Gaussian (left) and for the generalized
    Gaussian kernel function of degree 2 (right) ({\color{blue} $\bullet$}
    represents no hits).}
    \label{fig:kernelFunctionHitCountComparison}
\end{figure}

\figref{inferenceAlgoComparison} (bottom-right) shows the mean-hits number versus the performance for the Gaussian kernel and the generalized Gaussian kernel of degree 2.
It reaffirms that the performance depends on the number of hits rather than the number of particles, as noted previously. 
This explains the source of the speedup for the generalized Gaussian kernel, as the sharper extent reduces the number of hits.
See \figref{kernelFunctionHitCountComparison} for a visual plot.

\begin{table}[]
\caption{
Quality and speed tradeoffs for various particle kernel functions.}
\vspace{-3mm}
\footnotesize
\begin{center}
\begin{tabular}{l|ll|ll}
\toprule
 & \multicolumn{2}{c|}{\texttt{Tanks \& Temples}} & \multicolumn{2}{c}{\texttt{Deep Blending}} \\
 Particle\textbackslash{}Metric & \multicolumn{1}{c}{PSNR$\uparrow$} & \multicolumn{1}{c|}{FPS$\uparrow$} & \multicolumn{1}{c}{PSNR$\uparrow$} & \multicolumn{1}{c}{FPS$\uparrow$} \\ \midrule
Gaussian (reference) & 23.03  & 143 &  29.89  & 77 \\
Generalized Gaussian ($n=2$) & 22.68 & 277 & 29.74 & 141 \\
2D Gaussians & 22.70 & 241 & 29.74 & 122 \\
Cosine wave modulation & 22.77 & 268 & 29.79 & 126 \\
\bottomrule
\end{tabular}
\end{center}
\vspace*{3mm}
\label{tab:generalized_particles}
\end{table}

\begin{figure*}[!t]
    \centering
    \includegraphics[width=1\textwidth]{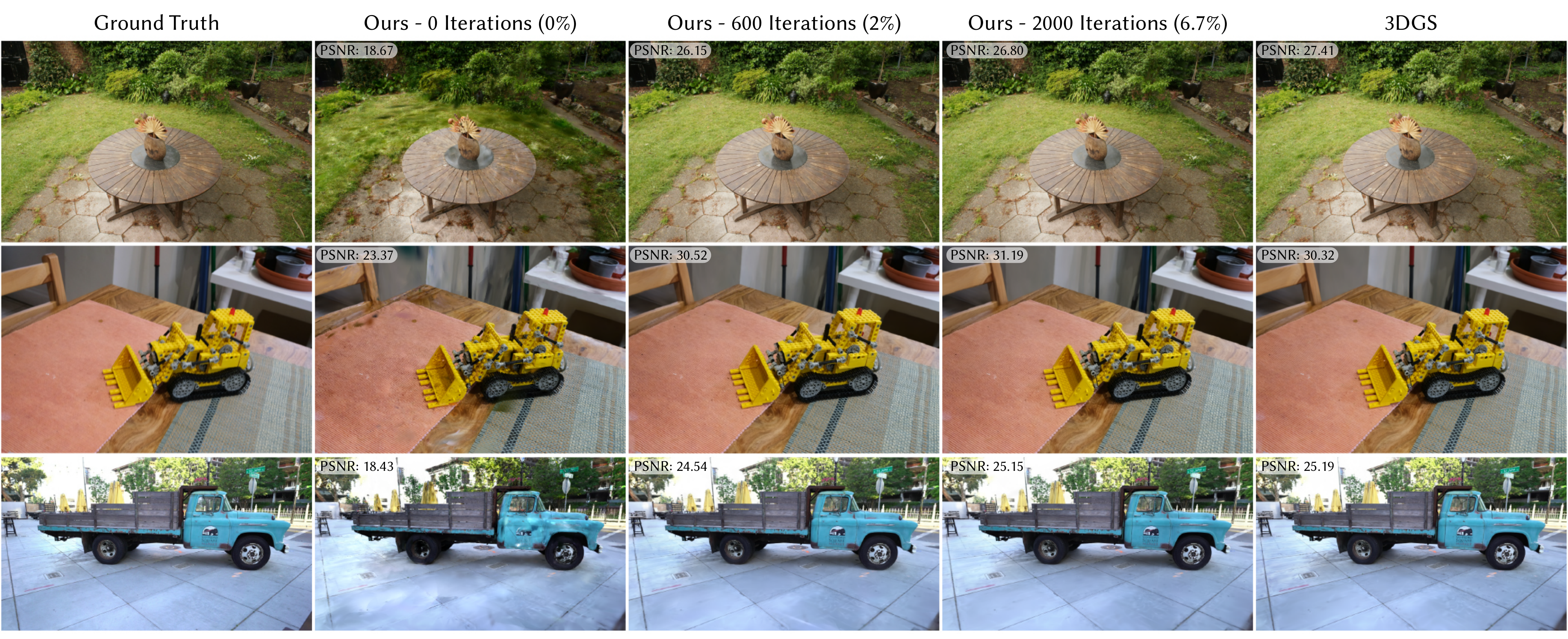}
    \vspace{-7mm}
    \caption{\textbf{3DGS Finetuning:} Qualitative results of finetuned models from pretrained 3DGS~\cite{kerbl3Dgaussians} checkpoints after different numbers of iterations.}
    \label{fig:finetuning_3dgs_qualitative}
\end{figure*}

\begin{figure}
    \centering
    \includegraphics[width=1\columnwidth]{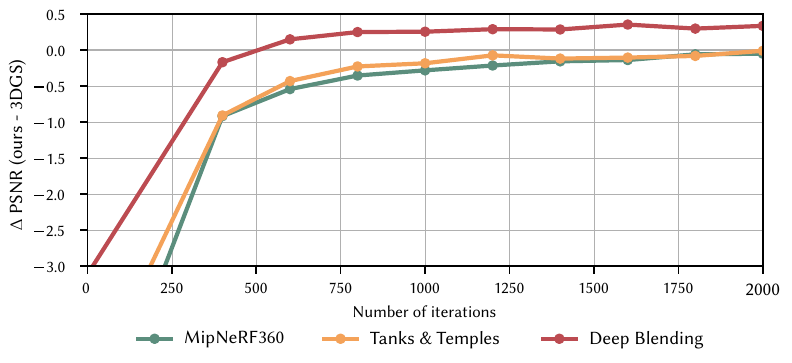}
    \vspace{-5mm}
    \caption{\textbf{Finetuning Pretrained 3DGS Models:} After only 500 iterations of finetuning, we can recover most of the perceptual quality of 3DGS~\cite{kerbl3Dgaussians}. After 2k iterations we match or outperform 3DGS across all datasets.}
    \label{fig:finetuning_3DGS}
\end{figure}

\subsubsection{Hit Buffer Size}

\figref{inferenceAlgoComparison} (top-right) measures the effect of the particle buffer size $\ParticleTraceBatchSize$,
which determines how many particle hits are gathered during each raycast before stopping to evaluate their response.
\emph{False rejected hits} are hits which are traversed, but not collected into the buffer because it is full with closer hits; these hits often must be re-traversed later.
\emph{False accepted hits} are hits which are gathered into the buffer, but ultimately do not contribute to radiance because the transmittance threshold is already met.
Both of these false hits harm performance, and choosing the particle batch size is a tradeoff between them.
We find $\ParticleTraceBatchSize=16$ to offer a good compromise and use it in all other experiments.

\vfill{}

\pagebreak
\section{Applications}
\label{sec:applications}

Most importantly, efficient differentiable ray tracing enables new applications and techniques to be applied to particle scene representations.

\subsection{Ray-Based Effects}
\label{sec:RayBasedEffects}

\begin{figure*}[!t]
    \centering
    \includegraphics[width=1\textwidth]{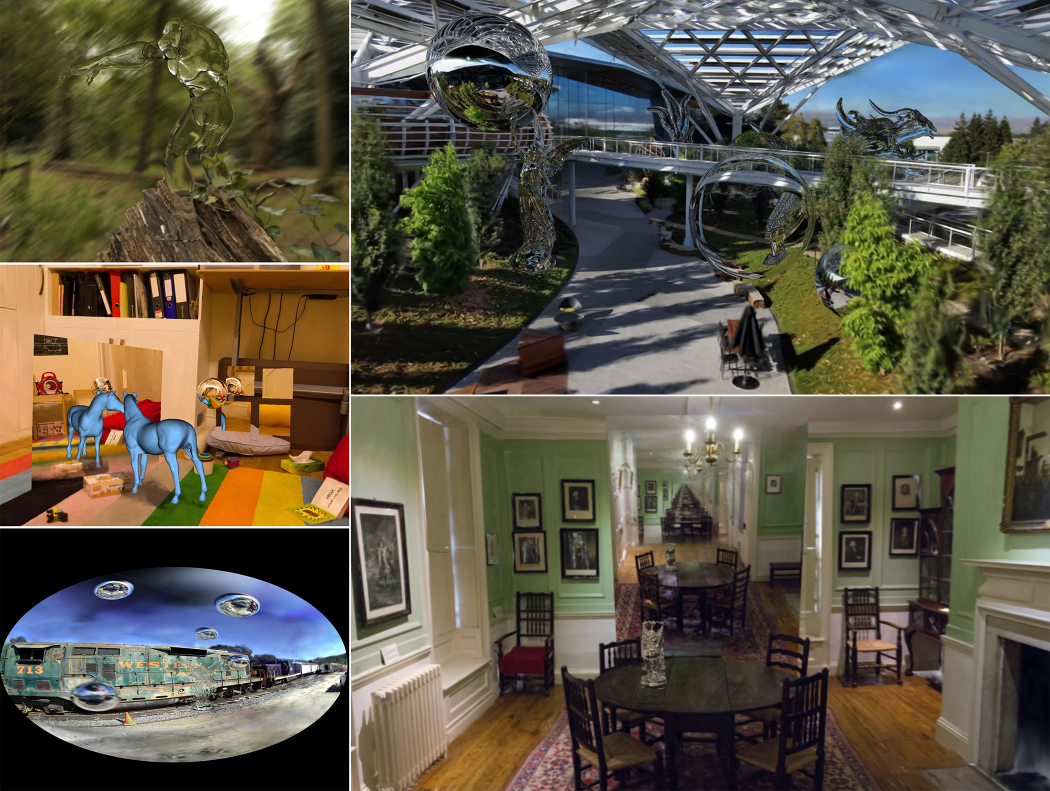}
    \caption{\textbf{Rendering Effects:} 
    The ray traced nature of our reconstructions allows seamless integration
    with traditional ray traced operations for reflecting and refracting rays,
    as well as casting shadows on nearby particles, as well as applying camera effects.}
    \label{fig:playground_gallery}
\end{figure*}

First, we the radiance field rendering pipeline with a variety of visual effects which are naturally compatible with ray tracing (\figref{playground_effects} and ~\ref{fig:playground_gallery}).
Here we demonstrate only manually-specified forward rendering, although inverse rendering in concert with these effects is indeed supported by our approach, and is a promising area for ongoing work.

\paragraph{Reflections, Refractions and Inserted Meshes}
Optical ray effects are supported by interleaved tracing of triangular faces and Gaussian particles.
Precisely, we maintain an extra acceleration structure consisting only of mesh faces for additional inserted geometry. 
When casting each ray in  \procref{raygen}, we first cast rays against inserted meshes; if a mesh is hit, we render all particles only up to the hit, and then compute a response based on the material.
For refractions and reflections, this means continuing tracing along a new redirected ray according to the laws of optics. %
For non-transparent diffused meshes, we compute the color and blend it with the current ray radiance, then terminate the ray.

\paragraph{Depth of Field}
Following \cite{mueller2022instant}, we simulate depth of field by progressively tracing multiple independent ray samples per pixel (spp), weighted together with a moving average to denoise the output image. The examples in Figures~ \ref{fig:playground_effects} and \ref{fig:playground_gallery} use 64-256 spp, although convergence is often reached with fewer samples by selecting subsamples with low discrepancy sequences \cite{Burley2020Scrambling}.

\paragraph{Artificial Shadows}
Even in radiance field scenes with baked-in lighting, simple shadow effects can be emulated by casting a ray towards a point or mesh emitter, and artificially darkening the contribution from that point if the light is not visible.
We adopt this approach, casting shadow rays after computing the directional contribution from each particle.

\subsection{Instancing}

\begin{figure}
    \centering
    \includegraphics[width=\linewidth]{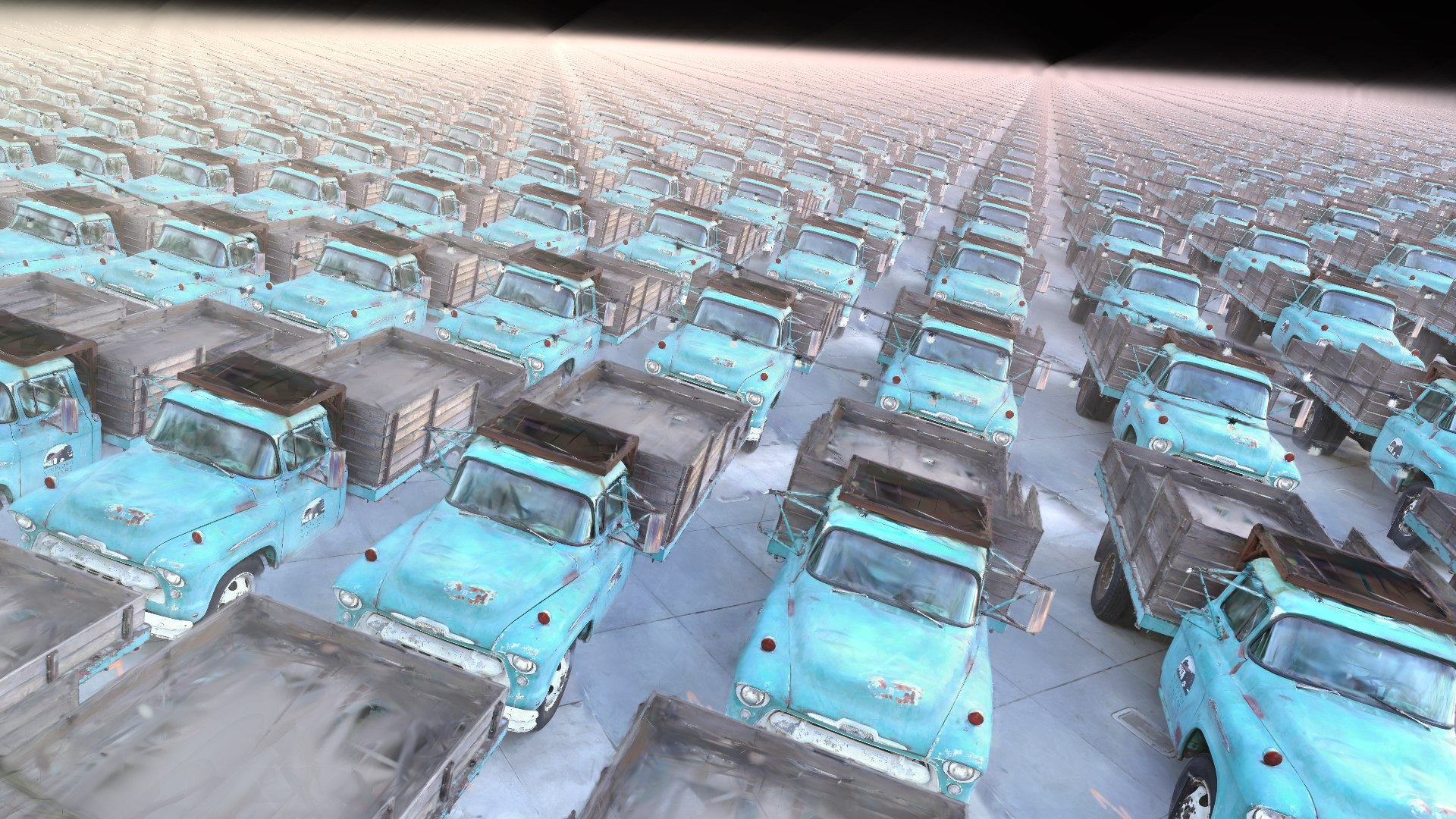}
    \caption{\textbf{Instancing:} 1024 instances of the \texttt{Tank \& Temples} \texttt{Truck}, rendered at more than 25 FPS.}
    \label{fig:instancing}
\end{figure}

In rendering, \emph{instancing} is a technique to render multiple transformed copies of an object with greatly reduced cost.
Although rendering libraries may support some form of instancing in the context of rasterization, the technique is particularly effective for ray tracing.
This is because repeated copies of an object can be stored as linked references in subtrees of the BVH, without actually duplicating the geometry.
This allows for scaling to 100s or 1000s of instanced objects at little additional cost---the same is not possible with rasterization.
Our efficient ray tracer enables particle scene data to be instanced, as shown in \figref{instancing} where we crop an object from a fitted scene and render 1024 copies of it at 25 FPS.

\begin{figure}
    \centering
    \includegraphics[width=\linewidth]{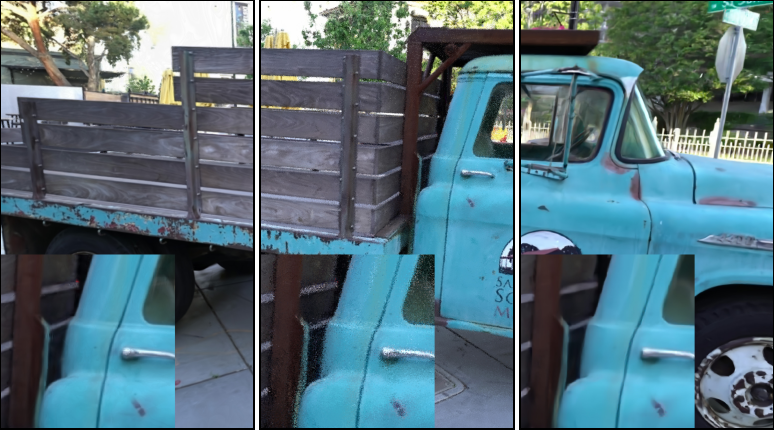}
    \caption{\textbf{Stochastic Sampling:} Left, scene rendered with our proposed algorithm. Center, rendered with stochastic sampling ($4$ samples). Right, denoised with the NVIDIA OptiX denoiser.}
    \label{fig:stochasticSamplingSample}
\end{figure}

\begin{figure*}
  \def\svgwidth{1.\linewidth}
  \centering
  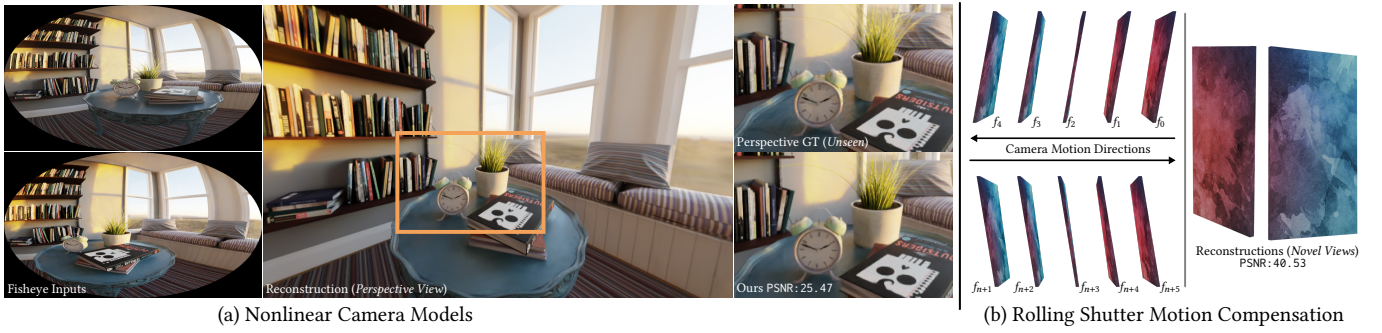
  \vspace{-7mm}
  \caption{
      {\bf Complex Cameras:} (a) Compared to rasterization-based approaches, our
      ray tracing-based formulation naturally supports complex camera models as
      inputs like distorted fisheye lenses (left), which can be re-rendered
      using different camera models like regular perspective cameras (right),
      achieving high reconstruction quality relative to \emph{unseen} references
      (see insets ({\color{palette2} $\bullet$}) - this \emph{synthetic} {\tt
      cozyroom} scene is by \cite{ma2021deblur}).
      (b) Ray tracing also naturally enables compensating for time-dependent
      effects like rolling shutter imagine sensors, which induce
      distortions due to sensor motion.
      This effect is illustrated on the left by multiple different frame tiles
      $f_i$ of a \emph{single} solid box rendered by a left- and right-panning
      rolling shutter camera with a top-to-bottom shutter direction.
      By incorporating time-dependent per-pixel poses in the reconstruction, our
      method faithfully recuperates the true undistorted geometry (right).
      }
  \label{fig:complex-cameras}
\end{figure*}

\subsection{Denoising and Stochastic Sampling}
\label{sec:denoising}

In ray tracing more broadly, research on stochastic integration techniques is key to highly sample-efficient yet perceptually compelling renders.
As a small demonstration, we show how our approach can be combined with stochastic sampling and denoising.

As discussed in \secref{tracing-algo}, our approach may be extended to stochastic-sampling by rejecting the hits received in the \emph{any-hit} program based on the importance sampling distribution $q=\hat{\ParticleResponse}(\bm{x})$. 
Since traversal stops as soon as the $k$-closest accepted samples are collected, this modification noticeably improves performance (see \figref{inferenceAlgoComparison} top-left).
This performance comes at a quality cost, but as shown in ~\figref{stochasticSamplingSample}, the resulting noise has statistics that an off-the-shelf denoiser can easily remove.

\subsection{Complex Cameras and Autonomous Vehicle Scenes}
\label{sec:RoboticsAndAutonomousVehicleScenes}

Ray tracing makes it easy, efficient, and accurate to render from exotic cameras which are far from ideal pinholes, such as highly-distorted fisheye cameras and those with rolling shutter effects (\figref{complex-cameras}).
While optical distortions for low-FOV lenses can be tackled to some extent by image rectification, and rolling shutter distortions can be approached by associating rasterized tiles to row/columns with consistent timestamps, both workarounds can't be applied simultaneously, as image rectification distorts the sets of concurrently measured pixels in a non-linear way.
In ray tracing, handling complex cameras simply means generating each ray with source and direction which actually correspond to the underlying camera, even if those rays may be incoherent and lack a shared origin.

\begin{figure}[h]
  \def\svgwidth{1.\linewidth}
  \centering
  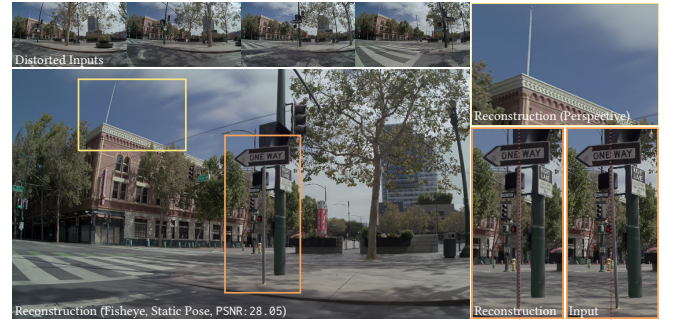
  \vspace{-6mm}
  \caption{
    {\bf AV Scene Reconstruction:} Real-world AV and robotics applications often
    have to respect both distorted intrinsic camera models and are, at the same
    time, affected by time-dependent effects like rolling shutter distortions as
    frames are exposed at high sensor speeds.
    Our ray tracing-based reconstruction is well-suited to handle \emph{both}
    challenges simultaneously, which we illustrated by an example of a side-facing
    top-to-bottom rolling shutter camera on an AV vehicle:
    the top inset ({\color{palette1} $\bullet$}) depicts faithful removal of the
    intrinsic camera model distortion by rendering with an undistorted camera,
    while the bottom inset ({\color{palette2} $\bullet$}) shows our ability to
    remove the apparent rolling-shutter distortions of the inputs by rendering
    from a \emph{static} camera pose (linear indicators ({\color{palette4}
    $\bullet$}) exemplify the complex distortion of the inputs).}
    \label{fig:rolling-shutter-av}
\end{figure}

\paragraph{Autonomous Vehicles}
The imaging systems used on autonomous vehicle (AV) platforms and other robot systems often incorporate such cameras, and it is very important to reconstruct and render them accurately in those applications.
\figref{rolling-shutter-av}, gives an example of an autonomous driving scene reconstructed from a side-mounted camera, which exhibits both apparent intrinsic camera and rolling shutter distortion effects. 

To further highlight the importance of accurately handling these effects, we
perform a quantitative evaluation of our method against its rasterization
equivalent 3DGS \cite{kerbl3Dgaussians} on autonomous driving scenes. 
We select 9 scenes from the \texttt{Waymo Open Perception} dataset~\cite{Sun2020CVPR_waymo_open} with no dynamic objects to ensure accurate reconstructions.
Both methods are trained with the images captured by the camera mounted on the front of the vehicle to reconstruct the scene.
We make several changes to the training protocol to adapt it to this data, including incorporating lidar depth, see the appendix for details.
For the case of 3DGS, we rectify the images and ignore the rolling shutter
effects, while with our tracing algorithm we can make use of the full camera
model and compensate for the rolling shutter effect (see
\figref{rolling-shutter-av}). 
Ray tracing achieves a rectified PSNR of $29.99$ on this benchmark, compared to $29.83$ for ordinary 3DGS---the improvement is modest, but it corresponds to correctly reconstructing important geometries, such as the signpost in~\figref{rolling-shutter-av}.

\section{Discussion}

\subsection{Differences Between Ray Tracing and Rasterization}
\label{sec:DifferencesRayTracingRasterization}
Here, we recap key differences between our ray tracer and the Gaussian splatting rasterizer proposed by ~\citet{kerbl3Dgaussians}.

\paragraph{Generality}
Splat rasterization accelerates rendering by processing a screen grid of pixel rays emanating from a single viewpoint in $16x16$ tiles, whereas ray tracing uses a BVH and can render along arbitrary distributions of rays from any direction.

\paragraph{Primary \vs{} General Rays}
As in graphics more broadly, ray tracing has significant benefits over rasterization to model general lighting and image formation.
In this work, we demonstrate a variety of effects such as reflection, refraction, depth of field, and artificial shadows enabled by ray tracing (\secref{RayBasedEffects}). In addition, we note that differentiable ray tracing opens the door to further research on global illumination, inverse lighting, and physical BSDFs.

\paragraph{Complex Cameras}
Using per-pixel rays, ray tracing can easily model more general image formation
processes that exhibit non-linear optical models such as highly-distorted and
high-FOV fisheye lenses, as well as time-dependent effects like rolling shutter
distortions, which originate from rows/columns exposed at different timestamps
(\cf~\cite{li2024usbnerf}).
These are important for robotics (\secref{RoboticsAndAutonomousVehicleScenes}), yet difficult or impossible to model with tile-based rasterization.

\paragraph{Speed}
\revAdded{
For forward rendering, our approach achieves real-time performance, and is only about $2\times$ slower than 3DGS's tiled rasterization in the basic case of rendering primary rays from pinhole cameras (see ~\tabref{fps_benchmark}).
For differentiable rendering to fit scenes, our ray tracing approach is $2\times$-$5\times$ slower than rasterization, mainly due to the need to rebuild the BVH (Section ~\ref{par:optimization}), and additional arithmetic needed to evaluate the backward pass as terms are no longer shared between pixels.
As an example, for the \emph{Tanks \& Temples}'s \emph{Truck} scene, ray tracing is $3.3\times$ slower per iteration of optimization, with a mean iteration time of 100ms (30ms spent on the BVH construction, 15ms on the forward pass and 30ms on the backward pass), while the rasterization requires 30ms per-iteration (7ms on the forward pass and 10ms on the backward pass). 
Furthermore, our approach enables training with incoherent rays as discussed in Section \ref{par:incoherentrays}, but in that case ray incoherency further increases the cost of ray tracing, leading to $5\times$ slower optimization.
}

\paragraph{Sub-Pixel Behavior}
Splat rasterization implicitly applies a pixel-scale convolution to Gaussians~\cite{zwicker2001surface}, whereas our ray tracer truly point-samples the rendering function and has no such automatic antialiasing. 
This may lead to differences of rendered appearance for subpixel-skinny Gaussian particles.
However, point-sampled rendering is well-suited to modern denoisers (see ~\secref{denoising}).

\paragraph{Interoperability}
It is possible to directly render scenes trained with the rasterizer under the ray tracing renderer, however due to 
subtle differences noted above, there will be a noticeable drop in quality when directly switching between renderers.
This can be quickly remedied with fine-tuning (see \figref{finetuning_3DGS}). 

\paragraph{Approximations}
Rasterization makes an approximation by evaluating directional appearance through the spherical harmonics $\ParticleHarmonics$ from a single ray direction, meaning each particle has constant appearance direction in all pixels of an image.
To support arbitrary distributions of rays in our ray tracer, each particle is evaluated exactly with the appropriate incoming ray direction.
Additionally, rasterization approximates depth ordering in 16x16 tiles.

\subsection{Limitations and Future Work}

Our ray tracer is carefully designed to make use of hardware acceleration and offers significant speedup over baseline implementations (\figref{inferenceAlgoComparison}), however ray tracing is still slower than rasterization when rendering from a pinhole camera.
Additionally, the need to regularly rebuild the BVH during training incurs additional cost and adds overhead to dynamic scenes.
Nonetheless, our implementation is still fast enough for training and interactive rendering, and more importantly it opens the door to many new capabilities such as distorted cameras and ray-based visual effects (\secref{applications}).
See \secref{DifferencesRayTracingRasterization} for an in-depth discussion of the trade-offs of rasterization \vs{} ray tracing in this setting.

Looking forward, this work creates great potential for further research on inverse rendering, relighting, and material decomposition on particle representations.
Indeed, recent work in this direction~\cite{R3DG2023, liang2023gs}\NS{we cite different works for this in the intro}, has relied on approximations due to the lack of an efficient ray tracer.
More broadly, there is much promising research to be done unifying advances in scene reconstruction from computer vision with the formulations for photorealistic rendering from computer graphics.

\begin{acks}
We are grateful to Hassan Abu Alhaija, Ronnie Sharif, Beau Perschall and Lars Fabiunke for assistance with assets, to Greg Muthler, Magnus Andersson, Maksim Eisenstein, Tanki Zhang, Dietger van Antwerpen and John Burgess performance feedback, to Thomas Müller, Merlin Nimier-David, and Carsten Kolve for inspiration, to Ziyu Chen, Clement Fuji-Tsang, Masha Shugrina, and George Kopanas for experiment assistance, and to Ramana Kiran and Shailesh Mishra for typo fixes. The manta ray image is courtesy of abby-design.
\end{acks}

{\small
\bibliographystyle{ACM-Reference-Format}
\bibliography{main}
}

\newpage
\clearpage

\newpage
\appendix

\section{Implementation and Training Details}
\label{sec:appendix_implementation_details}

We wrap the NVIDIA OptiX tracer as a Pytorch extension and train our representation using Adam optimizer for \num{30000} iterations. We set the learning rates for rotations, scales, and \revReplaced{albedo}{zeroth-order spherical harmonics} to \num{0.001}, \num{0.005}, and \num{0.0025}, respectively. The learning rate for the remaining spherical harmonics coefficients is $20$ times smaller than \revRemoved{that} for \revReplaced{albedo}{the zeroth-order coefficient}. Finally, we set the density learning rate to either \num{0.05} for high-quality settings or \num{0.09} for fast-inference settings. 

After initial \num{500} iterations, we start the densification and pruning process, which we perform until \num{15000} iterations are reached. To densify the particles, we accumulate 3D positional gradients, scaled by half the distance of each particle to the camera, to prevent under-densification in distant regions. In line with 3DGS~\cite{kerbl3Dgaussians}, we split the particles if their maximum scale is above $1\%$ of the scene extent and clone them otherwise. Pruning directly removes particles whose opacity is below \num{0.01}. Additionally, we employ a simple heuristic to cap the maximum number of particles to \num{3000000}. We denote this pruning strategy as \emph{visibility pruning}. Specifically, if the densification step results in more particles, we reduce their number to \num{2700000} by pruning particles with minimum accumulated weight contribution on the training views. Moreover, while densification and pruning are in effect and similar to 3DGS, we reset the particle densities to \num{0.01} every \num{3000} iterations. 
During training, we perform early stopping to terminate the tracing of rays whose accumulated transmittance falls below \num{0.001}. During inference, we increase this threshold to \num{0.03} for improved efficiency. We begin by solely training \revReplaced{albedo}{the constant spherical harmonic} and progressively increase the spherical harmonics' degree every \num{1000} iterations, up to a maximum of \num{3}. We update the BVH every iteration and reconstruct it after each pruning and densification.

For experiments with random-rays, during the last \num{15000} iterations, we sample random rays across all training views with a batch size of $2^{19}=\num{524288}$, and only use the $L1$ loss to supervise the particles.

\subsection{Autonomous Vehicles}

To fit autonomous vehicle scenes, we modify our training protocol, including incorporating lidar and depth supervision.
To initialize, we randomly sample 1 million lidar points visible in at least one training image.
These points are assigned an initial color via lookup projected into a training image, and assigned an initial scale based on the distance to the closest recorded camera pose.
During training, we use lidar to supervise depth; in our ray tracer depth is computed by integrating the distance along the ray to each sample as if it were radiance.
Note that in 3DGS, lidar depth must be approximated by projecting lidar rays onto camera images, yet in ray tracing lidar rays can be directly cast into the scene.
Additionally, we reconstruct the sky following~\cite{Rematas_2022_CVPR} and employ a directional MLP which predicts the color of the sky based on the ray direction.
A sky segmentation is included as input, and used to supervise ray opacities computed from the particle scene.

\vfill{}

\section{Additional Experiments and Ablations}
\label{sec:appendix_ablations}

\begin{table}[]
\caption{Quantitative evaluation on the \texttt{NeRF Synthetic} dataset~\cite{mildenhall2020nerf}}
\begin{center}
\resizebox{\columnwidth}{!}{
\begin{tabular}{l|S[table-format=3.2]S[table-format=3.2]S[table-format=3.2]S[table-format=3.2]S[table-format=3.2]S[table-format=3.2]S[table-format=3.2]S[table-format=3.2]S[table-format=3.2]}
\toprule
& \multicolumn{9}{c}{\texttt{NeRF Synthetic}} \\
Method &  \multicolumn{1}{c}{Chair} & \multicolumn{1}{c}{Drum} & \multicolumn{1}{c}{Ficus} & \multicolumn{1}{c}{Hotdog} & \multicolumn{1}{c}{Lego} & \multicolumn{1}{c}{Materials} & \multicolumn{1}{c}{Mic} & \multicolumn{1}{c}{Ship} & \multicolumn{1}{c}{Mean} \\
\midrule
NeRF & 33.00 & 25.01 & 30.13 & 36.18 & 32.54 & 29.62 & 32.91 & 28.65 & 31.10 \\
MipNeRF & 35.14 & 25.48 & 33.29 & 37.48 & 35.70 & 30.71 & 36.51 & 30.41 & 33.09 \\
INGP  & 35.00 & 26.02 & 33.51  & 37.40&  36.39 &29.78 & 36.22  &31.10 & 33.18\\   
AdaptiveShells  & 34.94 & 25.19& 33.63& 36.21& 33.49& 27.82& 33.91& 29.54& 31.84\\
\midrule
3DGS (paper) & 35.83 & 26.15& 34.87& 37.72& 35.78 & 30.00& 35.36&   30.80& 33.32\\
\midrule
Ours (reference)  & 35.90 & 25.77 & 35.94 & 37.51 & 36.01 & 29.95 & 35.66 & 30.71 & 33.48 \\

Ours & 36.02 & 25.89 & 36.08 & 37.63 & 36.20 & 30.17 & 34.27 & 30.77 & 33.38 \\
\bottomrule
\end{tabular}
}
\end{center}
\label{tab:nerf_synthetic}
\end{table}

\begin{figure*}[!t]
    \centering
    \includegraphics[width=1\textwidth]{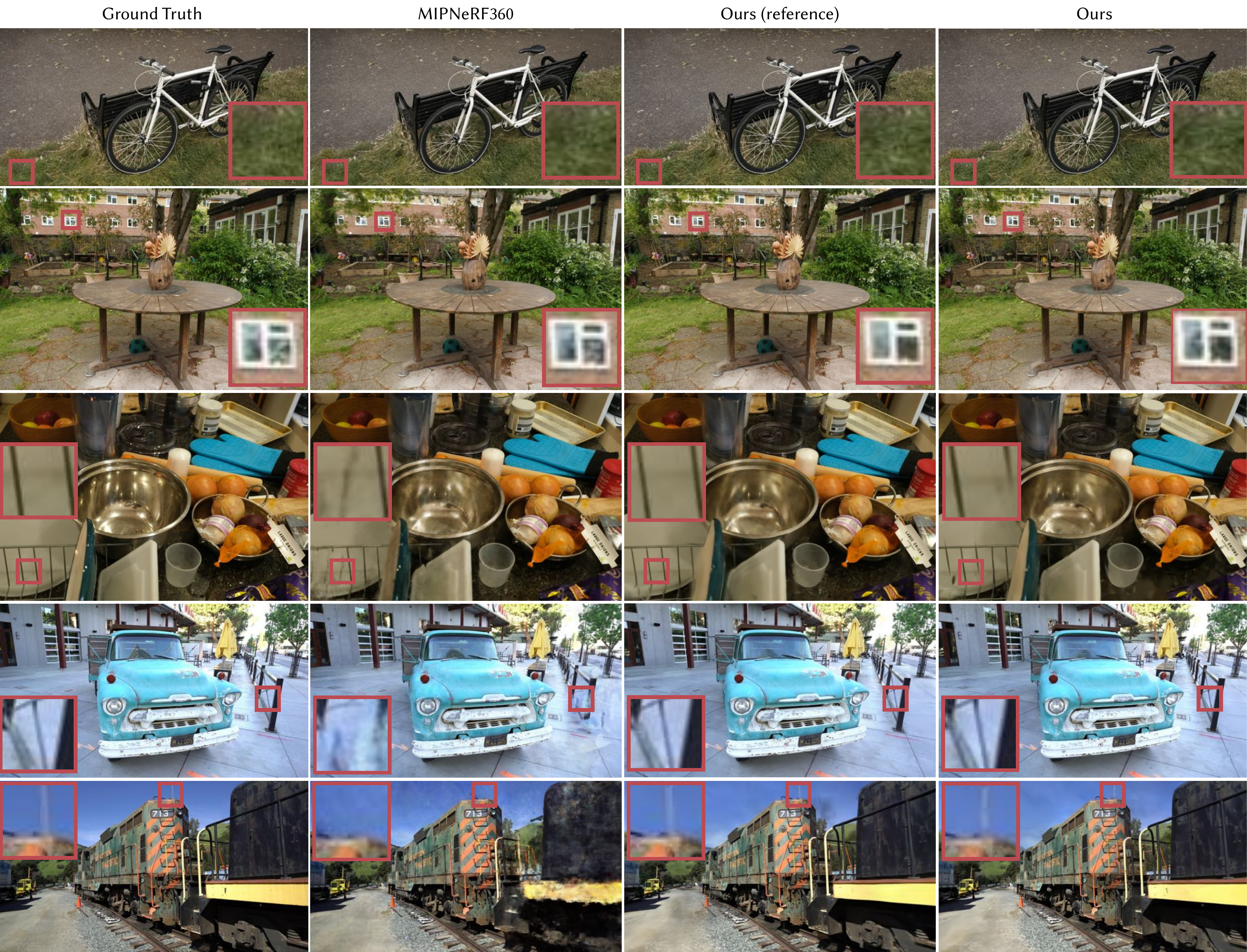}
    \caption{
    \textbf{Additional Qualitative Comparisons:} 
    novel-view synthesis results relative to the MIPNeRF360 baseline (insets
    ({\color{palette4} $\bullet$}) show per-result closeups).
    }
    \label{fig:appendix_results_gallery}
\end{figure*}

Figure~\ref{fig:appendix_results_gallery} shows qualitative comparisons of our method against MIPNeRF360~\cite{barron2022mipnerf360}. The zoomed-in insets demonstrate that both of our settings achieve comparable or better renderings with sharp features.
The first three rows contain scenes from the \texttt{MipNeRF360} dataset, while the last two rows feature scenes from \texttt{Tanks \& Temples}.

As mentioned in Section \ref{sec:appendix_implementation_details}, we propose a simple visibility pruning strategy to prevent the number of particles from exceeding a certain threshold. \tabref{num_gaussians_ablation} presents an ablation study on the maximum number of allowed particles for scenes in two datasets: \texttt{Tanks \& Temples} and \texttt{Deep Blending}. 
When densification causes the number of particles in the scene to exceed the threshold, we prune the least visible particles based on their accumulated contribution to the training views, reducing the number of particles to 90\% of the threshold.
The results show that our visibility pruning strategy, which filters out particles that contribute the least to the rendered views, maintains quality even with as few as one million particles.

\begin{table}[]
\caption{Quantitative PSNR ablation on the maximum number of allowed particles using \textit{ours}. }
\begin{center}
\resizebox{\columnwidth}{!}{
\begin{tabular}{l|S[table-format=3.2]S[table-format=3.2]S[table-format=3.2]S[table-format=3.2]S[table-format=3.2]S[table-format=3.2]}
\toprule
& \multicolumn{6}{c}{\emph{Maximum Allowed Particles}} \\
Dataset &  \multicolumn{1}{c}{$1\times10^6$} & \multicolumn{1}{c}{$2\times10^6$} & \multicolumn{1}{c}{$3\times10^6$} & \multicolumn{1}{c}{$4\times10^6$} & \multicolumn{1}{c}{$5\times10^6$} & \multicolumn{1}{c}{$6\times10^6$}\\
\midrule
\texttt{Tanks \& Temples} & 23.21&23.19&23.20&23.14&23.15&23.20 \\
\texttt{Deep Blending} &  29.24&29.17&29.23&29.14&29.24&29.15\\
\bottomrule
\end{tabular}
}
\end{center}
\label{tab:num_gaussians_ablation}
\end{table}

\end{document}